\documentclass[apj]{emulateapj}

\shorttitle{An extremely curved jet in PKS 2136+141}
\shortauthors{Savolainen et al.}

\begin{document}

\journalinfo{{\rm Accepted for publication in} The Astrophysical Journal.}
\submitted{Received 2005 October 19; accepted 2006 April 19}


\title{An Extremely Curved Relativistic Jet in PKS\,2136+141}


\author{Tuomas Savolainen, Kaj Wiik\altaffilmark{1}, Esko
Valtaoja\altaffilmark{2}} \affil{Tuorla Observatory, University of
Turku, V\"ais\"al\"antie 20, FI-21500 Piikki\"o, Finland}
\email{tukasa@utu.fi}

\author{Matthias Kadler\altaffilmark{3}, Eduardo Ros}
\affil{Max-Planck-Institut f\"ur Radioastronomie, Auf dem H\"ugel 69,
53121 Bonn, Germany}

\author{Merja Tornikoski} \affil{Mets\"ahovi Radio Observatory,
Helsinki University of Technology \\ Mets\"ahovintie 114, FI-02540
Kylm\"al\"a, Finland}

\and

\author{Margo F. Aller, Hugh D. Aller} \affil{Department of Astronomy,
University of Michigan \\ Dennison Building, Ann Arbor, MI 48109, USA}


\altaffiltext{1}{Institute for Space and Astronautical Science, VSOP-Group,
3-1-1, Yoshinodai, Sagamihara, Kanagawa 229-8510, Japan}
\altaffiltext{2}{Department of Physics, University of Turku, Finland}
\altaffiltext{3}{Radioastronomisches Institut der Universit\"at 
Bonn, Auf dem H\"ugel 71, 53121 Bonn, Germany}


\begin{abstract}
We report the discovery of an extremely curved jet in the radio-loud
quasar \objectname{PKS\,2136+141}.  Multi-frequency Very Long Baseline
Array (VLBA) images show a bending jet making a turn-around of 210
degrees in the plane of the sky, which is, to our knowledge, the
largest ever observed change in the position angle of an astrophysical
jet. Images taken at six different frequencies, from 2.3 to 43 GHz,
reveal a spiral-like trajectory, which is likely a sign of an
intrinsic helical geometry. A space-VLBI image, taken with the HALCA
satellite at 5 GHz and having comparable resolution to our
ground-based 15 GHz data, confirms that the bend is a
frequency-independent structure. VLBA monitoring data at 15 GHz,
covering eight years of observations, show knots in the jet clearly
deviating from ballistic motion, which suggests that the bending may
be caused by a growing helical Kelvin-Helmholtz normal mode. The jet
appearance suggests a helical wave at a frequency well below the
``resonant'' frequency of the jet, which indicates that the wave is
driven by a periodic perturbation at the base of the jet. We fit the
observed structure in the source with a helical twist, and we find
that a simple isothermal model with a constant wave speed and
wavelength gives a good fit.  The measured apparent velocities
indicate some degree of acceleration along the jet, which together
with an observed change in the apparent half-opening angle of the jet
allow us to estimate the changes in the angle between the local jet
direction and our line of sight. We suggest that the jet in
\objectname{PKS\,2136+141} is distorted by a helical Kelvin-Helmholtz
normal mode externally driven into the jet (e.g. by precession), and
that our line of sight falls within the opening angle of the helix
cone.
\end{abstract}




\section{INTRODUCTION}
A significant fraction of extragalactic jets show some degree of
bending -- from slightly curved jets up to a complete turn-around of
almost 180$\degr$. Recently, in their large study of jet kinematics of
radio-loud active galactic nuclei, \citet{kel04} measured vector
velocities for 60 bright jet features (also called components). They
found that approximately a third of these components show a
significant non-radial motion, i.e. the direction of their velocity
vector differs by at least $3\sigma$ from the mean structural position
angle of the jet. If these observed velocities trace the underlying
jet flow, their result indicates that bends in jet direction are very
common.

For core-dominated radio sources with high optical polarization, there
is a well-known bimodal distribution of the angles between jets in
parsec and kiloparsec scales, with a main peak of misalignment angles
around 0$\degr$ and a secondary peak around 90$\degr$
\citep{pea88,lis01}. However, a large-angle misalignment exceeding
120$\degr$ is rare \citep{wil86,tin98,lis01}. Up to today, the largest
observed $\Delta$P.A. is 177$\degr$ in the gamma-ray blazar
\objectname{PKS\,1510-089}, which shows a jet bending almost directly
across our line of sight \citep{hom02}. Since core-dominated radio
sources have jets oriented close to our line of sight, all intrinsic
variations in the jet trajectories are exaggerated in projection --
often to a large degree. This implies that rather small intrinsic
bends can manifest themselves as large-angle misalignments between the
jet axes observed on parsec and kiloparsec scales, or as high as
$\sim90\degr$ turns in the VLBI images.

Observations of relativistic jets in parsec scales provide evidence
that AGN jets can exhibit ``wiggling'' structures
(e.g. \objectname{4C\,73.18}, \citet{roo93}; \objectname{3C\,345},
\citet{zen95}; \objectname{3C\,273}, \citet{lob01};
\objectname{3C\,120}, \citet{har05}) reminiscent of helically twisted
patterns. It has been proposed that Doppler boosting together with
parsec-scale jets traveling in helical paths could explain the excess
of sources showing 90$\degr$ misalignment angle between pc and kpc
scale jets without invoking an uncomfortable 90$\degr$ intrinsic
curvature \citep{con93}. As the number of sources showing apparently
helical structures has grown, the helical jet models have become
increasingly popular also as an explanation for the (quasi)periodic
flux variations in AGN \citep{abr99,ost04}. However, the mechanism
producing an apparently helical shape of the jet is unclear -- as are
the explanations also for more modest observed bends.

The ``corkscrew'' structure of the jet in the well-known galactic
source \objectname{SS\,433} is successfully explained by ballistic
motion of material ejected from a precessing jet nozzle \citep{sti02},
and a similar model has been suggested also for several extragalactic
jets showing ``wiggling'' (e.g. \objectname{3C\,273}, \citet{abr99};
\objectname{BL\,Lac}, \citet{sti03}; \objectname{OJ\,287},
\citet{tat04}). In these models, the jet precession is either due to
the Lense-Thirring effect in a case of misalignment between the
angular momenta of accretion disk and a Kerr black hole (see
\citet{cap04} and references therein), or due to a binary black hole
system where a secondary black hole tidally induces the precession.

Contrary to the above-mentioned cases, \citet{lis03} report that
although the powerful radio source \objectname{4C\,+12.50} exhibits a
jet ridge line highly reminiscent of that in \objectname{SS\,433}, it
is most likely due to streaming instead of ballistic motion. Streaming
helical motion arises naturally from spatial stability analysis of
relativistic jets, since the jets are unstable against growing
Kelvin-Helmholtz normal modes \citep{har87}. Provided there is a
suitable perturbation mechanism present in the inner part of the jet,
the distortion waves propagating down the jet can displace the whole
jet (helical fundamental mode) or produce helically twisted patterns
on the jet surface (fluting modes). If the jet carries a large scale
electric current (so-called Poynting flux dominated jets), it is in
addition unstable against magnetic kink instability, which could also
produce observed ``wiggling'' structures \citep{nak04}.

Not all jets with observed bends exhibit ``wiggling'' structure, and
many of the observed changes in the jet direction can be explained
without invoking helical motions. Proposed explanations for curving
jets include ram pressure due to winds in the intracluster medium, a
density gradient in a transition to the intergalactic medium and
deflections by massive clouds in the interstellar medium. Most likely,
different mechanisms work in different sources. It would be valuable
to be able to reliably identify the reason for bending in individual
sources, since the observed properties of the bend -- correctly
interpreted -- can constrain several physical parameters of the jet
and the external medium (see e.g. \citet{har03} for the case of K-H
instabilities).

In this paper, we present Very Long Baseline Array (VLBA) images from
a dedicated multi-frequency observation and from the VLBA 2 cm
Survey\footnote{http://www.cv.nrao.edu/2cmsurvey/} \citep{kel98}
showing that \objectname{PKS\,2136+141} (\objectname{OX\,161}), a
radio-loud quasar at moderately high redshift of 2.427, has a
parsec-scale jet, which appears to bend {\it over} 180$\degr$ on the
plane of the sky, being -- to our knowledge -- the largest ever
observed change in the position angle of an astrophysical jet (other
sources showing very pronounced changes in the jet direction include
e.g. \objectname{PKS\,1510-089} \citep{hom02}, \objectname{1803+784}
\citep{bri99}, and \objectname{NRAO\,150} (I. Agudo et al., in
preparation)). In VLA images, \objectname{PKS\,2136+141} is a compact
source, showing no extended emission on arcsecond scales
\citep{mur93}. Both 5 GHz \citep{fom00} and 15 GHz \citep{kel98} VLBA
observations reveal a core-dominated source with a short, slightly
bending jet.

\begin{figure*}
\epsscale{0.95}
\plotone{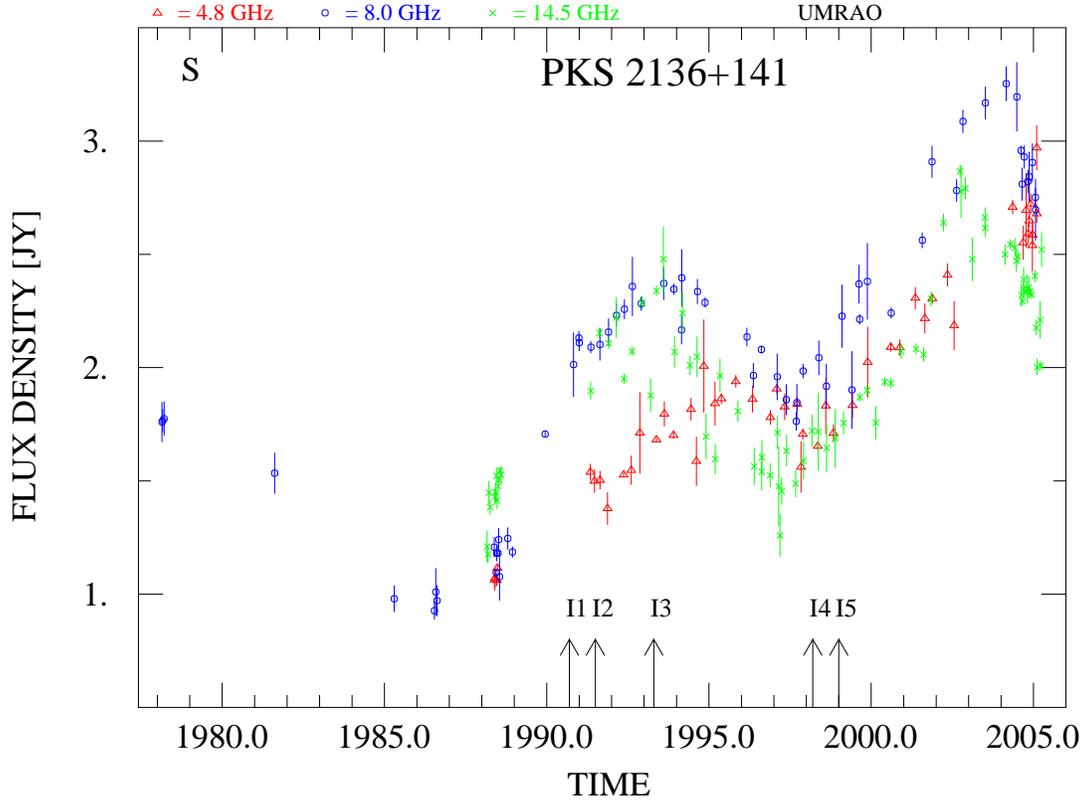}
\caption{Weekly averages of total flux density for PKS\,2136+141
(OX\,161) from UMRAO centimetre band monitoring. Observations at 14.5,
8.0 and 4.8 GHz are denoted by (green) crosses, (blue) circles and
(red) triangles, respectively. Extrapolated ejection epochs of VLBI
components are marked with arrows. Component I6 is left out from the
figure because it is seen only at the last epoch of the VLBA 2\,cm
Survey monitoring data, and therefore, we cannot accurately determine
its ejection epoch. The ejection of I6 must have occured at sometime
between the ejection of I5 and April 2004, when it is first
observed. See the on-line edition for a color version of the figure.}
\label{umrao}
\epsscale{1.0}
\end{figure*}

Originally, \citet{tor01} identified \objectname{PKS\,2136+141} as a
candidate gigahertz-peaked spectrum source (GPS) having a slightly
inverted spectrum up to 8-10 GHz in the intermediate-to-quiescent
state and a clearly inverted spectrum ($\alpha \ge +0.5$) during
outbursts. The high turnover frequency reported in \citet{tor01} even
puts \objectname{PKS\,2136+141} in the class of high frequency peakers
(HFP). Although \citet{tor05} have recently classified the source as a
flat spectrum radio source having a convex spectrum only during
outbursts, the simultaneous continuum spectra from RATAN-600
\citep[S. Trushkin, private communication]{kov99} do show a convex
shape also in the intermediate-to-quiescent state, albeit with a
little lower peak frequency. The source is variable at radio
frequencies showing a factor of $\sim 3$ variations in cm-wavelength
flux curves with characteristic time scale of $\sim 5-6$ years (see
Figure~\ref{umrao}). The last strong outburst started around 1998 and
peaked in late 2002 and in early 2004 at 14.5 and 8 GHz,
respectively. This indicates that both, our multi-frequency
observations in 2001 and 15 GHz VLBA monitoring during 1995-2004,
caught the source during a major flare.

The paper is organized as follows: the multi-frequency VLBA data
demonstrating the 210$\degr$ bend of the jet, together with a
space-VLBI observation from the HALCA satellite, are presented in
\S~2. In \S~3, we present a kinematic analysis, derived from over
eight years of the VLBA 2 cm Survey monitoring data at 15 GHz,
indicating non-ballistic motion of the jet components. Also, changes
in $\beta_{\mathrm{app}}$ and in the apparent half-opening angle of
the jet are investigated. In \S~4, possible reasons for observed
bending are discussed and a helical streaming model explaining the
observed structure is presented. Conclusions are summarized in
\S~5. Throughout the paper we use a contemporary cosmology with $H_0$
= 71\,km\,s$^{-1}$\,Mpc$^{-1}$, $\Omega_{M}$ = 0.27 and $\Omega
_\Lambda$ = 0.73. For this cosmology and a redshift of 2.427, an
angular distance of 1 mas transforms to 8.2 pc and a proper motion of
0.1 mas yr$^{-1}$ to an apparent speed of $9.2\,c$. We choose the
positive spectral index convention, $S_{\nu} \propto \nu^{+\alpha}$.

\begin{deluxetable*}{ccccccc}
\centering
\tablewidth{0pt}
\tablecaption{Parameters of the Images for Figures ~\ref{mutka} and ~\ref{halcamap}
 \label{mapar}}
\tablecolumns{7}
\tablehead{\colhead{Frequency} &
\colhead{$\Theta_{b,\textrm{\scriptsize{maj}}}$} &
\colhead{$\Theta_{b,\textrm{\scriptsize{min}}}$} & \colhead{P.A.} &
\colhead{rms noise} & \colhead{Peak Intensity} & \colhead{Contour
$c_0$\tablenotemark{a}} \\ \colhead{(GHz)} & \colhead{(mas)} &
\colhead{(mas)} & \colhead{(deg)} & \colhead{(mJy beam$^{-1}$)} &
\colhead{(mJy beam$^{-1}$)} & \colhead{(mJy beam$^{-1}$)}}
\startdata
2 & 6.12 & 3.02 & -11.7 & 0.4 & 1153 & 1.2 \\
5 & 3.37 & 1.70 & -12.5 & 0.2 & 1811 & 0.6 \\
5\tablenotemark{b} & 0.75 & 0.59 & -26.6 & 8.3 & 708 & 24.9 \\
8 & 1.98 & 1.01 & -7.8  & 0.2 & 1890 & 0.6 \\
15 & 0.90 & 0.50 & -7.5 & 0.5 & 1429 & 1.5 \\
22 & 0.63 & 0.36 & -7.9 & 1.2 & 1153 & 3.6 \\
43 & 0.43 & 0.18 & -16.1 & 1.5 & 512 & 4.5 \\
\enddata

\tablenotetext{a}{Contour levels are represented by geometric series
$c_0(1,...,2^n)$, where $c_0$ is the lowest contour level indicated in
the table (3x rms noise).}

\tablenotetext{b}{HALCA image}
\end{deluxetable*}

\section{MULTI-FREQUENCY OBSERVATIONS \label{multi}}
On May 2001 we made multi-frequency polarimetric VLBI observations of
four HFP quasars, including \objectname{PKS\,2136+141}, using the
VLBA. Observations were split into a high frequency part (15, 22 and
43 GHz), which was observed on the 12th of May, and into a low
frequency part (2.3, 5 and 8.4 GHz) observed on the 14th of May. Dual
polarization was recorded at all frequencies.

\begin{figure*}
\epsscale{1.09}
\plotone{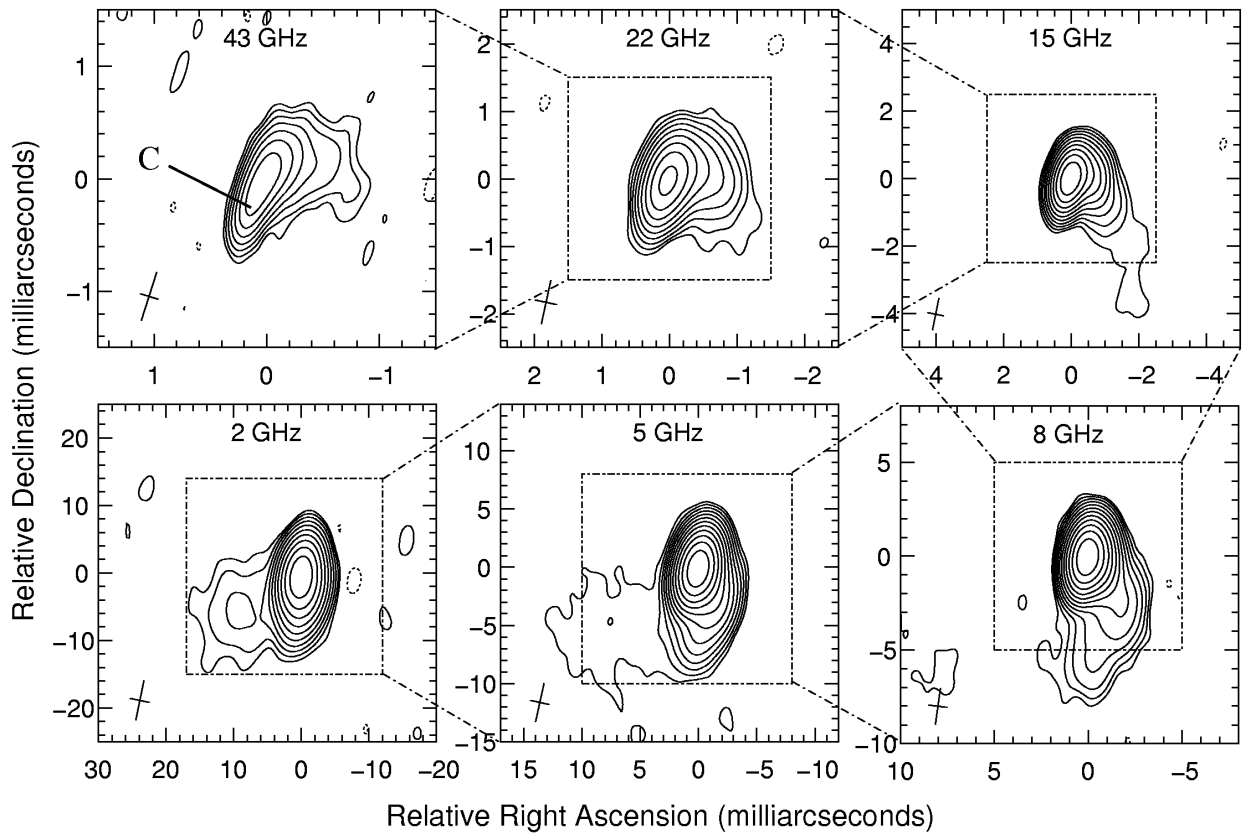}
\caption{Total intensity VLBA observations of PKS\,2136+141 from May
2001. The compiled figure shows images of the source at all six
observed frequencies (2.3, 5, 8.4, 15, 22, and 43 GHz). The letter
``C'' in the top left panel marks the location of the core. The size
and orientation of the beam is shown in the lower left corner of each
image. Peak intensities and contour levels are given in
Table~\ref{mapar}.}
\label{mutka}
\epsscale{1.00}
\end{figure*}

\subsection{Reduction of the VLBA Data}
The data were correlated with the VLBA correlator in Socorro and were
postprocessed in Tuorla Observatory using the NRAO's Astronomical
Image Processing System, AIPS, \citep{bri94,gre88} and the Caltech
{\sc difmap} package \citep{she97}. Standard methods for VLBI data
reduction and imaging were used. An {\it a priori} amplitude
calibration was performed using measured system temperatures and gain
curves. For the high frequency data (15--43 GHz), a correction for
atmospheric opacity was applied. After the removal of a parallactic
angle phase, a single-band delay and phase off-sets were calculated
manually by fringe fitting a short scan of data of a bright source. We
did manual phase calibration instead of using pulse-cal tones, because
there were unexpected jumps in the phases of the pulse-cal tones
during the observations. Global fringe fitting was performed, and the
delay difference between right- and left-hand systems was removed (for
the purpose of future polarization studies). Bandpass corrections were
determined and applied before averaging across the channels, after
which the data were imported into {\sc difmap}.

In {\sc difmap} the data were first phase self-calibrated using a point
source model and then averaged in time. We performed data editing in a
station-based manner and ran several iterations of {\sc clean} and
phase self-calibration in Stokes I. After a reasonable fit to the
closure phases was obtained, we also performed amplitude
self-calibration, first with a solution interval corresponding to the
whole observation length. Solution interval was gradually shortened as
the model improved by further cleaning. Final images were produced
with the Perl library
FITSPlot\footnote{\url{http://personal.denison.edu/~homand/}}.

We have checked the absolute flux calibration by comparing the
extrapolated zero baseline flux density of our compact calibrator
source \objectname{1749+096} at 5, 8.4, and 15 GHz to the single-dish
measurements made at University of Michigan Radio Astronomy
Observatory (UMRAO), and at 22 and 43 GHz to the fluxes from
Mets\"ahovi Radio Observatory's quasar monitoring program at 22 and 37
GHz, respectively \citep{ter04}. The flux densities agree to 5\% at
8.4, 15, 22 and 37/43 GHz, and to 8\% at 5 GHz, which is better than
the expected nominal accuracy of 10\% for the {\it a priori} amplitude
calibration. Being unable to make a flux check for the 2.3 GHz data,
we conservatively estimate it to have an absolute flux calibration
accurate to 10\%.

In order to estimate the parameters of the emission regions in the
jet, we model-fitted to the self-calibrated (\textit{u,v}) data in
{\sc difmap}. The data were fitted with a combination of elliptical
and circular Gaussian components, and we sought to obtain the best
possible fit to the visibilities and to the closure phases. Several
starting points were tried in order to avoid a local minimum fit. We
note that since the source structure is complex, the models are not
unique, but rather show one consistent parameterization of the
data. Based on the experiences in error estimation reported by several
groups, we assume uncertainties in component flux $\sim 5$\%, in
position $\sim 1/5$ of the beam size (or of the component size if it
is larger than the beam), and in size $\sim 10$\% (see
e.g. \citet{jor05} and \citet{sav05} for recent discussions on the
model fitting errors). Although \citet{jor05} use larger positional
uncertainties for weak knots having flux densities below 50 mJy, we
use $\sim 1/5$ of the beam size (or of the component size) also for
these components.  Bigger uncertainties would result in such a large
ratio of the individual errors to the scatter of the component
positions (about the best-fit polynomial describing the component
motion) that it would be statistically unlikely (see
\S~\ref{velocities}).
 
A detailed description of the polarization data reduction and imaging
in Stokes Q and U together with polarization images will appear in
T. Savolainen et al. (in preparation).

\subsection{Reduction of the HALCA data}

\begin{figure}
\includegraphics[width=\columnwidth]{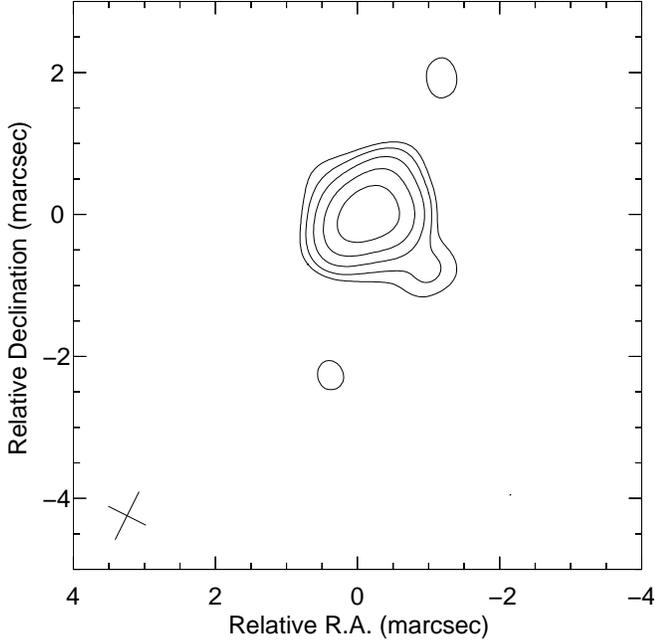}
\caption{5 GHz space-VLBI image of PKS\,2136+141 observed
  on 1998 May 28. Beam size, peak intensity and contour levels are
  given in Table~\ref{mapar}.}
\label{halcamap}
\end{figure}

The 5 GHz space-VLBI observations of \objectname{PKS\,2136+141} were
carried out as a part of the VSOP survey program \citep[R. Dodson et
al. in preparation]{hir00} on 1998 May 28.  In addition to the HALCA
satellite, the array consisted of Arecibo, Haartebesthoek, and the
Green Bank (140 ft) telescopes. The data were correlated using the
Penticton correlator and reduced using the same standard methods as
for the ground based images. Due to the small aperture of the HALCA
satellite, its weight was increased to 10 to persuade the fitting
algorithm to take better account of the long space baselines.

The dynamic range of the resulting image is rather small. This is due
to a component that had a strong effect only to a single baseline
between Arecibo and Green Bank. Because the (\textit{u,v}) coverage is
very limited, all attempts to model this component diverged. By
manually changing the model to follow the visibility amplitudes in
this scan, we estimate that the position of this component is about 7
mas at a PA of 100$^\circ$.  Because this component could not be
formally included in the model, the residuals are rather strong and
thus the imaging noise is high.

\subsection{Source Structure from the Multi-Frequency Data \label{structure}}
Figure~\ref{mutka} displays images of \objectname{PKS\,2136+141} at
all six observed frequencies. In the images at 15, 22 and 43 GHz,
uniformly weighted (\textit{u,v})-grids are employed in order to achieve
the best possible resolution, whereas normal weighting is used in the
low frequency maps to highlight diffuse, low surface brightness
emission. The restoring beam sizes, peak intensities, off-source rms
noise, and contour levels of the images are given in
Table~\ref{mapar}.

The multi-frequency images strikingly reveal a jet which gradually
bends $210\degr$ with its structural P.A. turning clockwise from
-27$\degr$ at 43 GHz to +123$\degr$ at 2.3 GHz. We identify the bright
and the most compact model component lying in the south-east end of
the jet in the 15--43 GHz images as the core, and mark it with a letter
``C''. The identification is confirmed by a self-absorbed spectrum of
the component, as will be shown later. Between about 0.4-1.0 mas from
the core, the jet turns over $90\degr$, which is visible in 15--43 GHz
images, and in the image taken at 8.4 GHz, the jet direction continues
to turn clockwise $\sim 50\degr$ at about 3.5 mas from the core. There
is also evident bending in the 5 and 8.4 GHz images: a curve of $\sim
70\degr$ takes place at about 6 mas south of the core. It is not
totally clear, whether the trajectory of the jet is composed of a few
distinct bends or whether it is a continuous helix. However, the
gradual clockwise turn and the apparent spiral-like appearance of the
jet in the multi-frequency images are highly reminiscent of a helical
trajectory. Also, it seems unlikely that the jet goes through at least
three consecutive deflections with all of them having the same sense
of rotation in the plane of the sky.

The 5 GHz space-VLBI image of \objectname{PKS\,2136+141} from 1998 May
28, i.e. three years before the multi-frequency VLBA observations,
shows a rather compact core-jet structure with an extended emission to
north-west (Figure~\ref{halcamap}). The jet takes then a sharp
$90\degr$ bend to south-west within about 1 mas from the core. This 5
GHz image shows a very similar curved structure near the core that can
be seen in our ground-based 15 GHz image with matching
resolution. Hence, the observed large-angle bending between the images
taken at different frequencies cannot be attributed to frequency
dependent opacity effects.

\begin{figure}
\includegraphics[width=\columnwidth]{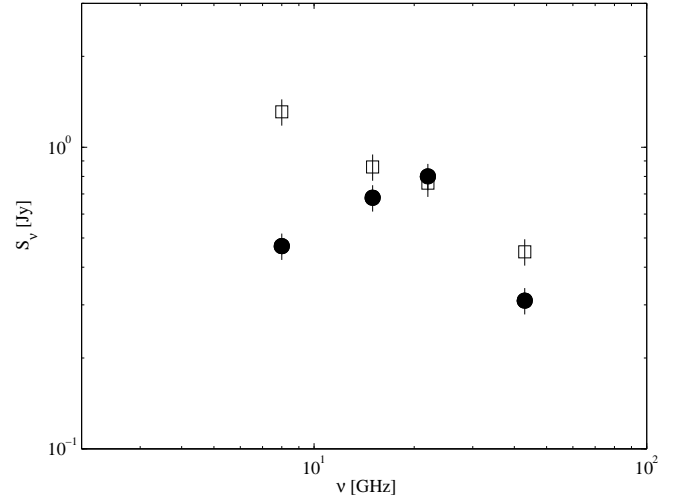}
\caption{Spectra of the core (filled circles) and the component I5
  (open squares) at 8--43 GHz on May 2001.} 
\label{spectrum}
\end{figure}

\objectname{PKS\,2136+141} is rather compact at all frequencies with a
maximum jet extent of approximately 15 mas corresponding to $\sim120$
pc at the source distance. The two brightest model components are the
core and a newly ejected component I5 (see \S 3 for component
identifications), which is located at $\sim 0.25$ mas from the core --
near the beginning of the first strong bend in the jet.
Figure~\ref{spectrum} shows the 8--43 GHz spectra of these two
components. At 2.3 and 5 GHz the angular resolution is too poor to
separate the core from I5, and hence, corresponding flux values are
omitted in the figure. The core has a synchrotron peak frequency of
$\sim 20$ GHz, while the newly ejected component I5 shows an optically
thin synchrotron spectrum with a spectral index $\alpha=-0.6$ and a
peak frequency below 8 GHz. The component I5 is brighter than the core at
every frequency except at 22 GHz, where almost equal flux densities
are measured. An optically thick spectrum of the core at frequencies
below 22 GHz indicates that I5 is brighter than the core also at 2.3
and 5 GHz, although the spectra cannot be measured at those
frequencies. The self-absorbed spectrum shown in Figure~\ref{spectrum}
confirms that we have correctly identified the core.

\section{VLBA 2\,CM SURVEY AND MOJAVE DATA \label{kin}}

\begin{figure*}
\plotone{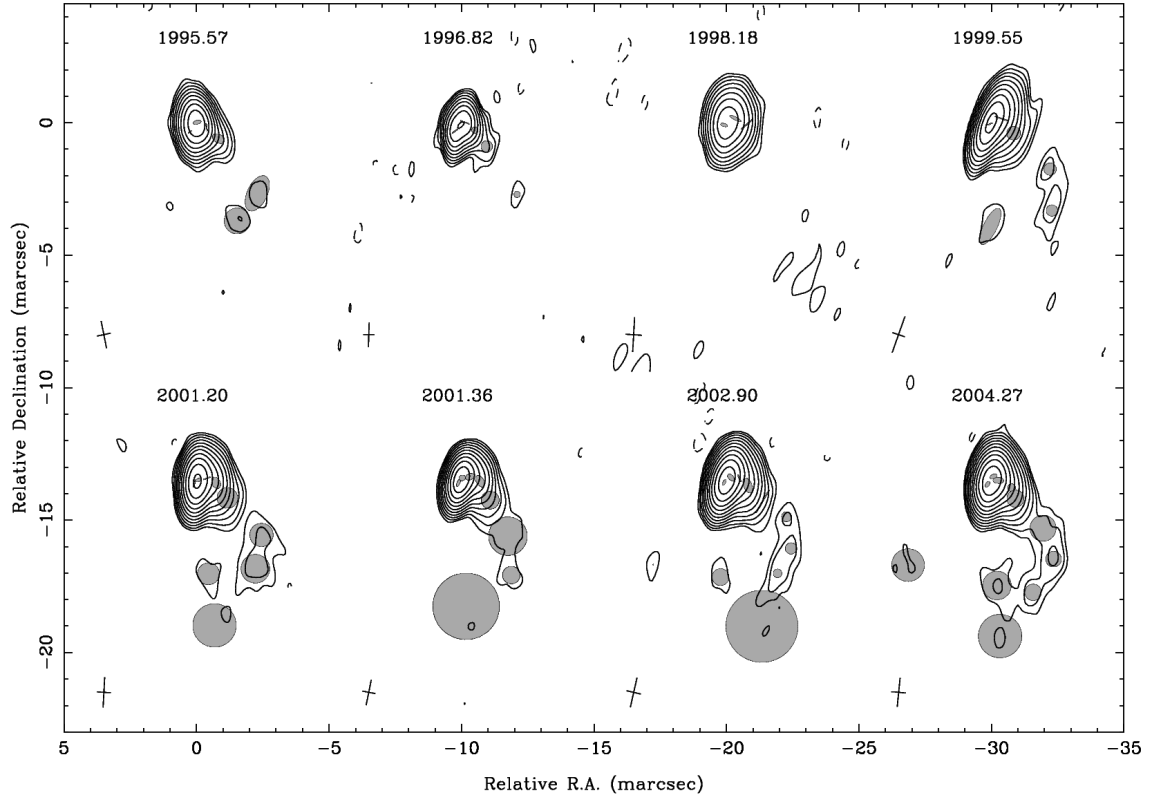}
\caption{VLBA images of PKS\,2136+141 at 15 GHz. The figure includes
observations from the VLBA 2\,cm Survey, from the MOJAVE Survey and a
15 GHz image from our multi-frequency data set observed in 2001.36. In
all images, a uniformly weighted (\textit{u,v})-grid is used. The
Gaussian components fitted to the visibility data are shown as
ellipses overlaid each image. The size and orientation of the beam is
shown in the lower left corner of each image. Peak intensities and
contour levels are given in Table~\ref{mapar2}.}
\label{2cm_data}
\end{figure*}

\begin{deluxetable*}{ccccccc}
\centering
\tablewidth{0pt}
\tablecaption{Parameters of the Images for Figure~\ref{2cm_data} \label{mapar2}}
\tablecolumns{7}
\tablehead{\colhead{Epoch} &
\colhead{$\Theta_{b,\textrm{\scriptsize{maj}}}$} &
\colhead{$\Theta_{b,\textrm{\scriptsize{min}}}$} & \colhead{P.A.} &
\colhead{rms noise} & \colhead{Peak Intensity} & \colhead{Contour
$c_0$\tablenotemark{a}} \\ \colhead{(yr)} & \colhead{(mas)} &
\colhead{(mas)} & \colhead{(deg)} & \colhead{(mJy beam$^{-1}$)} &
\colhead{(mJy beam$^{-1}$)} & \colhead{(mJy beam$^{-1}$)}}
\startdata
1995.57 & 1.01 & 0.53 & 11.5 & 0.9 & 1254 & 2.7 \\
1996.82 & 0.92 & 0.41 & -1.6 & 0.8 & 633 & 2.4 \\
1998.18 & 1.31 & 0.61 & -3.4 & 1.4 & 929 & 4.2 \\
1999.55 & 1.45 & 0.49 & -19.6 & 0.7 & 1429 & 2.1 \\
2001.20 & 1.14 & 0.52 & -3.1 & 0.4 & 1153 & 1.2 \\
2001.36 & 0.90 & 0.50 & -7.5 & 0.5 & 1429 & 1.5 \\
2002.90 & 1.15 & 0.55 & -12.9 & 0.4 & 2120 & 1.2 \\
2004.27 & 1.09 & 0.54 & -5.7 & 0.3 & 1745 & 0.9 \\
\enddata

\tablenotetext{a}{Contour levels are represented by geometric series
$c_0(1,...,2^n)$, where $c_0$ is the lowest \\ contour level indicated in
the table (3x rms noise).}

\end{deluxetable*}

Seven observations of \objectname{PKS\,2136+141} have been performed
as part of the VLBA 2\,cm Survey \citep{kel98,zen02} between 1995 and
2004 of which the last one (epoch 2004.27)\footnote{After
2004.27, two further MOJAVE observations of \objectname{PKS\,2136+141}
have been conducted.  The monitoring will be continued through 2006
and later.}  was part of the follow-up program MOJAVE \citep{lis05}.
Details on the survey observing strategy, the observing details and
the data reduction can be found in the mentioned publications.

{\sc clean} VLBI images of \objectname{PKS\,2136+141} have been
produced by applying standard self-calibration procedures with {\sc
difmap} for all seven epochs.  The calibrated visibility data were
fitted in the (\textit{u,v}) domain with two-dimensional elliptical Gaussian
components (see Figure~\ref{2cm_data}). Because of the complicated
source structure, no adequate model representation of the source could
be established with a smaller number of components than shown in
Figure~\ref{2cm_data}.  Moreover, the components are found to be
located close-by along the curved inner part of the jet, in many cases
separated from each other by considerably less than one beam
size. Special care had to be taken to find a model of this complicated
structure in a consistent way for all epochs.

\subsection{Source Kinematics}

\begin{figure}
\includegraphics[width=\columnwidth]{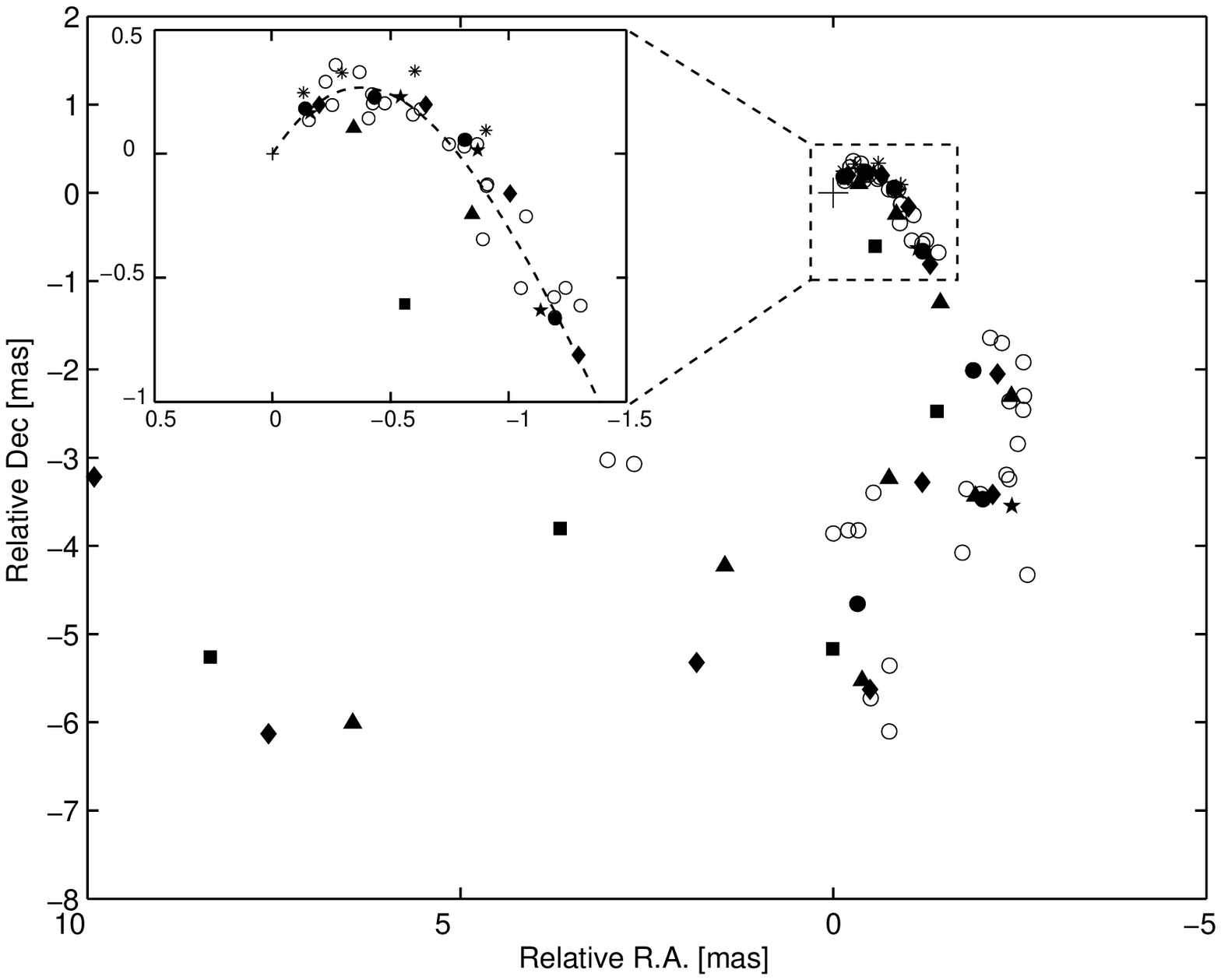}
\caption{Model-fit component locations. {\it Open circles} are model
components from the VLBA 2\,cm Survey monitoring and filled symbols show the
multi-frequency data: {\it squares} -- 2.3 GHz, {\it triangles} -- 5 GHz,
{\it diamonds} -- 8.4 GHz, {\it circles} -- 15 GHz, {\it stars} -- 22
GHz, and {\it asterisks} -- 43 GHz. The position of the core is marked
with a cross. The inset is an enlarged plot of the innermost 2 x 1.5 mas 
and it also displays an approximate ridge line of the jet.} 
\label{allcomp}
\end{figure}

A crucial question raised by the apparent helical shape of the jet in
Figure~\ref{mutka} is, whether it represents streaming motion or
whether the helix is due to a precessing jet nozzle ejecting material
that moves along ballistic trajectories. We have plotted positions of
all model-fit components from the multi-frequency data and from eight
years of the VLBA 2\,cm Survey monitoring data into
Figure~\ref{allcomp}. The open circles, which correspond to the
multiepoch data from the VLBA 2\,cm Survey, form a dense and strongly
curved region near the base of the jet, a $\sim 2$ mas long continuous
and slightly curved section 3.5 mas south-west of the core, and a
few small isolated groups. This subtle finding alone, without invoking
any component identification scenarios, demonstrates that
ballistic-motion models are unlikely to yield a meaningful
representation of the jet kinematics in
\objectname{PKS\,2136+141}. Brightening at certain points of the jet
can be either due to an increased Doppler factor (if a section of the
jet bends towards our line of sight), or due to an impulsive particle
acceleration in a standing shock wave (e.g. forming in the
bend). There seems to be a zone of avoidance in the 15 GHz data
between the base of the jet and the large western group, further
supporting this idea.

We have analyzed the jet kinematics using source models derived from
the VLBA 2\,cm Survey monitoring data. In a complicated source like
\objectname{PKS\,2136+141} it is often difficult to identify
components across epochs with confidence. We have based our
identifications on the most consistent trajectories, and on the flux
density evolution of the components. We restrict our analysis to a
full kinematical model for the inner 2 mas of the jet, since beyond
that distance a fully self-consistent model could not be established
due to the lower surface brightness and high complexity in the outer
region.

\subsubsection{Component Trajectories and Flux Evolution \label{sec_traj}} 

\begin{figure*}
\epsscale{0.95}
\plotone{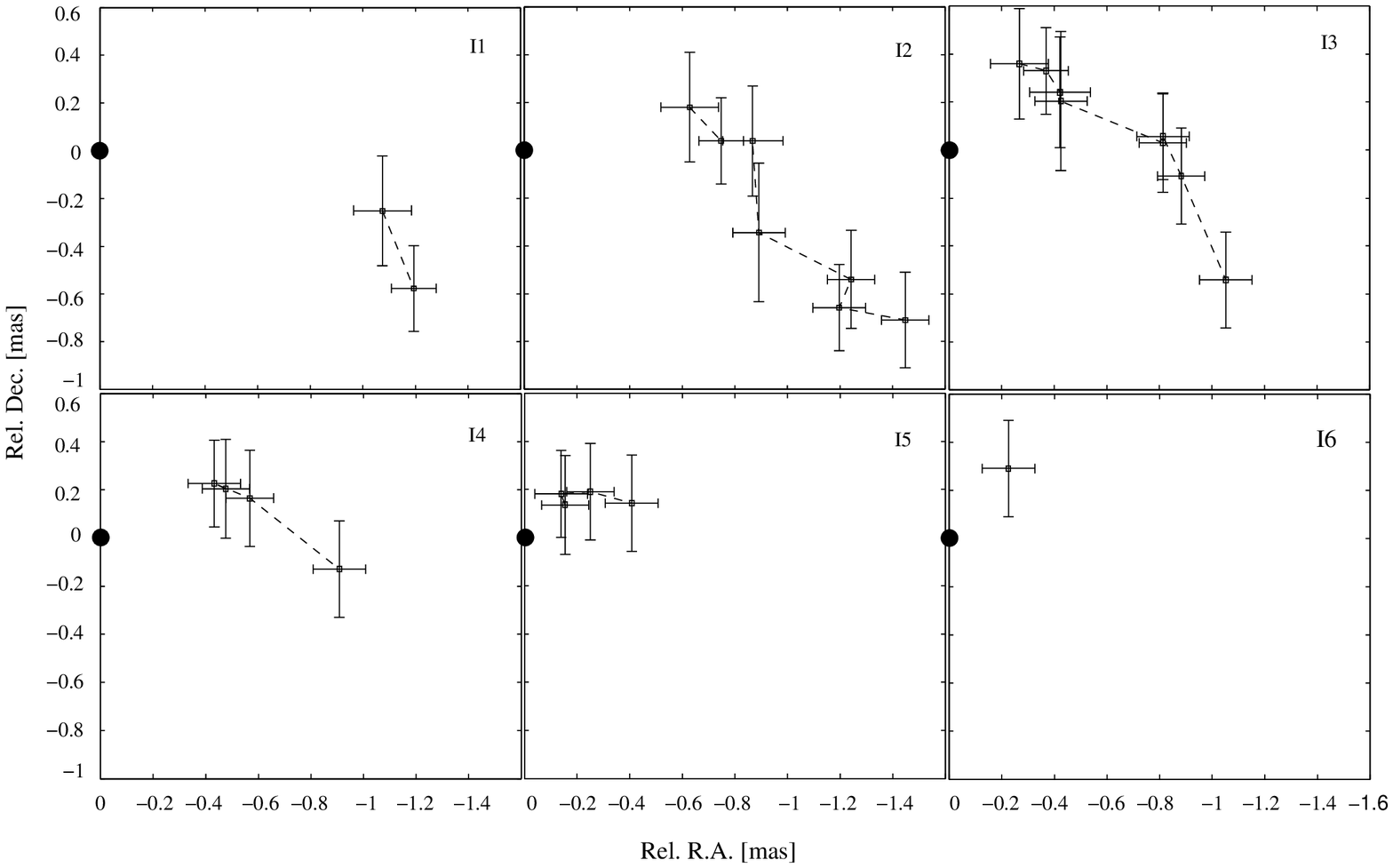}
\caption{Trajectories of components I1--I6. Positions are measured
relative to the core, which is placed into the origin in each
panel ({\it filled circle}). The errorbars correspond to $1/5$ of the beam size.}
\label{trajectories}
\epsscale{1.00}
\end{figure*}

\begin{figure}
\includegraphics[width=\columnwidth]{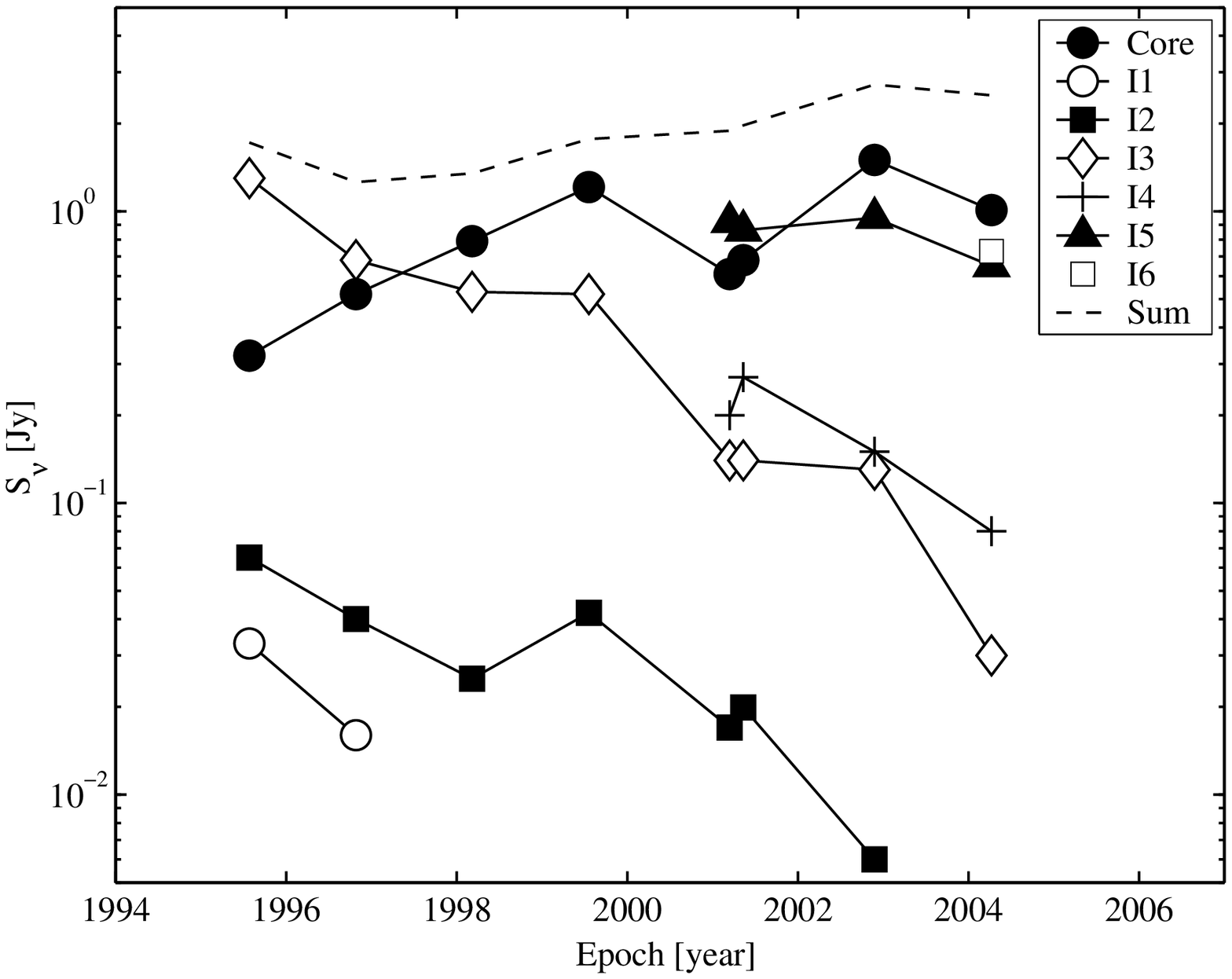}
\caption{Flux densities of the identified components from the VLBA 2\,cm
Survey monitoring.} 
\label{fluxes}
\end{figure}

Figure~\ref{trajectories} shows the trajectories of the six compact
components (I1-I6) which we have identified in the inner 2 mas of the
jet. The components travel towards south-west and their trajectories
do not extrapolate back to the core implying that the components
travel along a curved path, and the jet is non-ballistic. The
components fade below the detection limit within $\sim 1.6$ mas from
the core, and we cannot follow them further.

We present the flux density evolution of the six inner-jet components
in Figure~\ref{fluxes}. The overall picture is that the fluxes
decrease as the components move forward along their path. They reach
about 5--15 mJy by the time they have traveled to a distance of $\sim
1.5$ mas from the core, after which they are not seen anymore. The
knot I3 is the brightest component of the source at the first two
epochs, and its flux density evolution matches well with the strong
1993 flare visible in the 14.5 GHz UMRAO flux density curve (see
Figure~\ref{umrao}). The ejections of the components I1 and I2 (see
Table~\ref{speeds}) take place during the rising phase of the 1993
outburst, and the component I3 is ejected just before the outburst
reaches its maximum.

The core has a steadily rising flux density until 2001.20, when the
flux suddenly drops and two new components, I4 and I5, appear. At the
next two epochs, the core flux increases again and at the last epoch
(2004.27) there is a drop accompanied by an appearance of a new
component, I6. The brightening of the core during our monitoring and
ejections of components I4, I5, and I6 correspond to a strong total
flux density flare peaking in late 2002 at 14.5 GHz (again, see the
total flux density curve in Figure~\ref{umrao}).

The above-mentioned flux density evolution of the components is
self-consistent and it is well in accordance with the general behavior
of flat-spectrum radio quasars during strong total flux density flares
\citep{sav02}, i.e. a compact VLBI core is mostly responsible for the
rising part of the observed flares in single-dish flux curves, and a
new component appears into the jet during or after the flare peaks
accompanied by a simultaneous decrease in the core flux density. The
fact that the overall flux density evolution of the identified
components in \objectname{PKS\,2136+141} seems to obey the common
behavior identified for a number of other sources supports our
kinematical model. An intriguing detail is the consecutive ejection
of, not one, but three new components in connection with a single,
strong total flux density flare. This may indicate that there is some
substructure in the flares, i.e. the outbursts in 1993 and 2002 could
be composed of smaller flares. On the other hand, it could also mean
that a single strong event in the total flux density curve is able to
produce complicated structural changes in the jet, e.g. forward and
reverse shocks could both be visible, or there could be trailing
shocks forming in the wake of the main perturbance as theoretically
predicted by \citet{agu01} and recently observed in several objects,
most prominently in \objectname{3C\,111} \citep[Kadler et al. in
preparation]{jor05,kad05} and in \objectname{3C\,120} \citep{gom01}.

\subsubsection{Apparent Velocities and Acceleration \label{velocities}}

\begin{figure}
\includegraphics[width=\columnwidth]{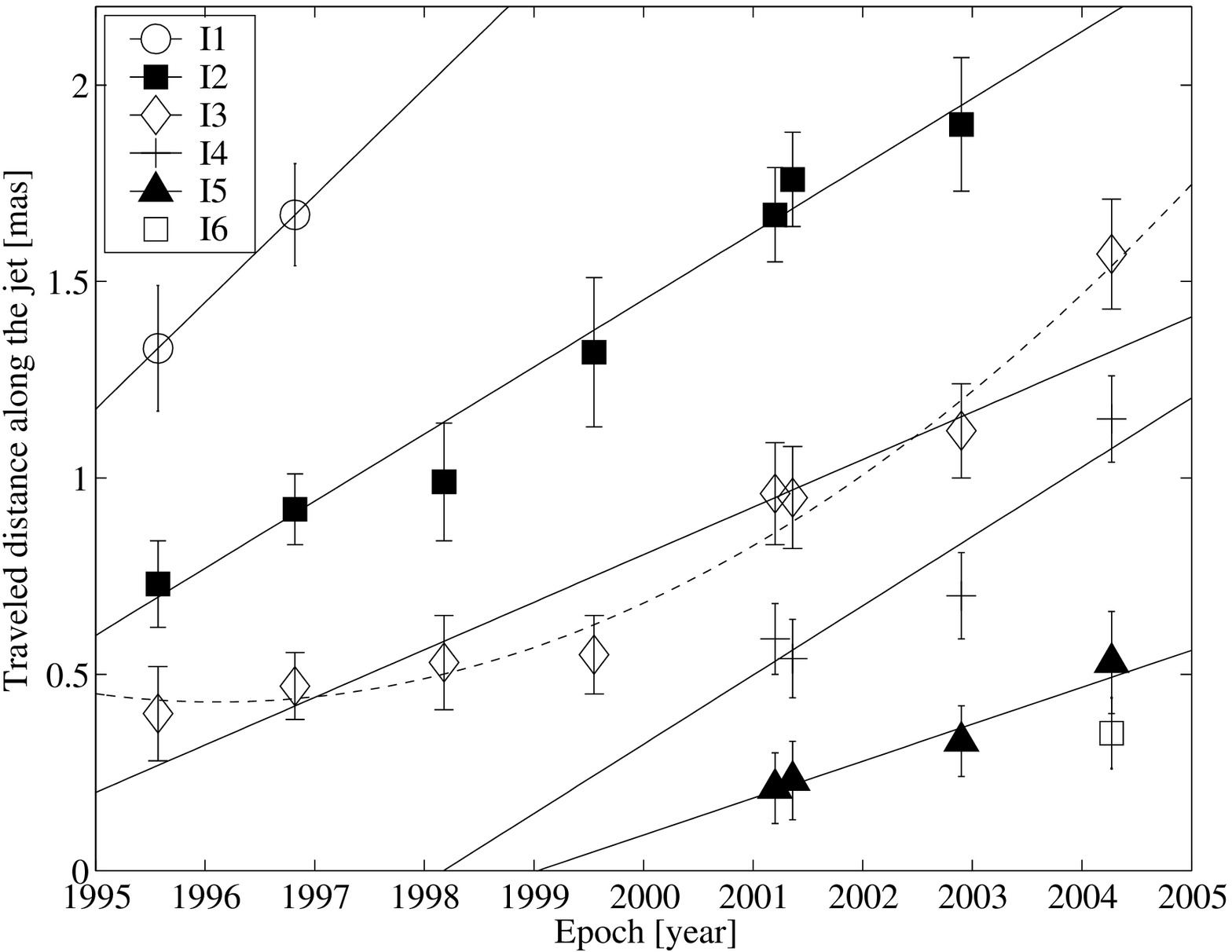}
\caption{Component motion along the jet ridge line. The symbols for different
component are: {\it open circles} -- I1, {\it filled squares} -- I2,
{\it open diamonds} -- I3, {\it crosses} -- I4, {\it filled
triangles} -- I5, and an {\it open square} -- I6. The {\it solid lines}
represent linear fits to the data, and the {\it dashed line}
represents a second-order fit.}  
\label{motion}
\end{figure}

Identification of the model-fit components across the epochs allows us
to estimate the velocities of the components and to search for
possible acceleration or deceleration. Since the components follow a
highly curved path, we choose to measure the distance traveled along
the jet by using an average ridge line, which is shown by the dashed
line in Figure~\ref{allcomp}. The ridge line is determined by fitting
a smooth function to the positions of model components within the
inner 2 mas of the jet.

Figure~\ref{motion} displays the motion of the components along the
jet. To measure the component velocities and to estimate the ejection
epochs, we have made standard linear least-squares-fits to the motions:
\begin{eqnarray}
l(t_i) & = & a_0+a_1(t_i-t_{\mathrm{mid}}), 
\end{eqnarray}
where $l$ is the traveled distance along the jet, $t_i$ is the epoch
of observation $(i=1,\ldots,N)$ and $t_{\mathrm{mid}}=(t_1 +
t_N)/2$. In order to search for a possible acceleration or a deceleration,
we have also fitted the components I2 and I3, which are observed over most
epochs, by second-order polynomials:
\begin{eqnarray}
l(t_i) & = & a_0+a_1(t_i-t_{\mathrm{mid}})+a_2(t_i-t_{\mathrm{mid}})^2.
\end{eqnarray}
To assess the goodness of the fits, we have calculated a {\it
relative} $\chi^2$ value for each fit. The $\chi^2$ values are
relative, not absolute, since we do not {\it strictly} know the real
$1\sigma$ measurement errors of the component positions but have
instead used $1/5$ of the beam size projected onto the ridge line.
Although the ``$1/5$ of the beam size'' estimate for the errors is
rather well-justified in the literature, it is likely to be
conservative. Thus the tabulated values of $\chi^2$ distribution do
not provide a good statistical test of the goodness of the fits in our
case. However, we assume that the beam sizes give a good estimate of
the relative errors between the epochs, and hence, only one scaling
factor (which is close to unity) for the positional errors (and for
the $\chi^2$ values) remains unknown. With this assumption, the
relative $\chi^2$ values can be used to compare the goodness of the
fits against each other. The parameters of the fitted polynomials are
given in Table~\ref{poly} together with relative reduced $\chi^2$
values and the number of the degrees of freedom, $\nu$, for each fit.

We have gathered the average angular velocities along the jet,
$\langle \mu \rangle = a_1$, the average apparent speeds, $\langle
\beta_\mathrm{app} \rangle$, and the epochs of zero separation, $T_0$,
for each component in Table~\ref{speeds}. For the second-order fits,
also the angular acceleration along the jet, $\dot{\mu} = 2 a_2$, is
reported.  The proper motions range from 0.09 to 0.27 mas yr$^{-1}$,
corresponding to the apparent superluminal velocities of
$8.7-25.1\,c$.

\begin{deluxetable}{ccccccc}
\tablewidth{0pt}
\tablecaption{Best-Fit Polynomials for the Motion of Components I1--I5
\label{poly}}
\tablecolumns{7}
\tablehead{\colhead{Comp.} & \colhead{$k$\tablenotemark{a}} &
\colhead{$a_0$} & \colhead{$a_1$} & \colhead{$a_2$} &
\colhead{$\chi^2 / \nu$\tablenotemark{b}} &
\colhead{$\nu$\tablenotemark{c}}}
\startdata
I1 & 1 & $1.50\pm0.10$ & $0.27\pm0.17$ & \nodata & \nodata & 0 \\
I2 & 1 & $1.32\pm0.05$ & $0.17\pm0.02$ & \nodata & 0.34 & 5 \\
   & 2 & $1.29\pm0.08$ & $0.17\pm0.02$ & $0.01\pm0.01$ & 0.37 & 4 \\
I3 & 1 & $0.80\pm0.04$ & $0.12\pm0.02$ & \nodata & 1.50 & 6 \\
   & 2 & $0.68\pm0.06$ & $0.13\pm0.02$ & $0.02\pm0.01$ & 0.43 & 5 \\
I4 & 1 & $0.81\pm0.05$ & $0.18\pm0.04$ & \nodata & 1.19 & 2 \\
I5 & 1 & $0.35\pm0.06$ & $0.09\pm0.04$ & \nodata & 0.12 & 2 \\
\enddata
\tablenotetext{a}{Order of the fitted polynomial.}
\tablenotetext{b}{The $\chi^2$ value is \textit{relative}; see text.}
\tablenotetext{c}{Number of the degrees of freedom in the fit.}
\end{deluxetable}

While the component I2 is well-fitted by a straight line, the linear
proper motion model does not seem to adequately represent the motion
of the component I3 as can be seen from Figure~\ref{motion}. On the
other hand, a second-order polynomial gives an acceptable fit to I3,
and its relative $\chi^{2}$ value divided by the number of the degrees
of freedom is by a factor of 3.5 smaller than the value for the
first-order polynomial. However, the small number of data points and
the uncertainty about the real $(1\sigma)$ errors in the component
positions makes it difficult to give a statistical significance of the
deviation from the constant speed model. We have used two approaches
to assess the significance. The most robust and the most
straightforward test is to use a statistic $F_{\nu_1,\nu_2} \equiv
(\chi^2_1/\nu_1)/(\chi^2_2/\nu_2)$, which has an $F$-distribution with
$(\nu_1,\nu_2)$ degrees of freedom, and which is independent of the
unknown scaling factor in the positional errors. This statistics can
be used to test whether the squared residuals of the linear model are
significantly larger than those of the accelerating model. As already
mentioned, for the first and the second-order polynomials in the case
of I3, $F_{6,5}=3.5$, which implies a difference in the residuals only
at the confidence level of $\alpha=0.10$; i.e. according to the
$F$-test, the statistical significance of the difference in the
residuals is only marginal. Another approach to this problem is to try
to estimate the real $1\sigma$ uncertainties of the component
positions by applying a method described by \citet{hom01}. They
estimated the uncertainties of the fitted parameters of their proper
motion models by using the variance about the best-fit model, and as a
by-product they obtained an upper bound estimate of the component
position uncertainty. Since the relative $\chi^2$ values divided by
the degrees of freedom are significantly less than unity for
components I2 and I5 (the situation of component I4 is discussed
later), we suspect that the real $1\sigma$ positional uncertainties
are in fact smaller than $1/5$ of the beam size. If the beam sizes
give a good estimate of the relative errors between the epochs, we can
try to estimate the uniform scaling factor for the positional
uncertainties by requiring that $\chi^2 \simeq \nu$ for the linear
proper motion models of I2, I4, and I5. The sum
$\Sigma_i(\chi_i^2/k^2-\nu_i)^2,\,(i=\mathrm{I2,I4,I5})$ is minimized
for the positional error scaling factor $k=0.77$; i.e. the $1\sigma$
positional uncertainties are $0.77 \times 1/5$ of the beam size
$\approx 0.08 - 0.15$ mas. If these $(1\sigma)$ uncertainties are
assumed, the linear fit to the motion of component I3 has
$\chi^2=15.2$ with 6 degrees of freedom, implying that the speed of I3
is {\it not} constant at the significance level of $\alpha=0.025$.

\begin{deluxetable}{cccccc}
\tablewidth{0pt}
\tablecaption{Proper Motions of Model-Fit Components \label{speeds}}
\tablecolumns{6}
\tablehead{\colhead{Comp.} & \colhead{$k$\tablenotemark{a}} &
\colhead{$\dot{\mu}$} &\colhead{$\langle \mu \rangle$} & \colhead{$\langle
\beta_{\mathrm{app}} \rangle$} & \colhead{$T_{0}$} \\ \colhead{} &
\colhead{} & \colhead{(mas yr$^{-2}$)} & \colhead{(mas yr$^{-1}$)}&
\colhead{$(c)$} & \colhead{(yr)}}
\startdata
I1 & 1 & \nodata & $0.27\pm0.17$ & $25.1\pm15.2$ & $1990.7\pm3.0$ \\
I2 & 1 & \nodata & $0.17\pm0.02$ & $15.8\pm1.8$ & $1991.5\pm0.6$ \\
   & 2 & $0.01\pm0.02$ & $0.17\pm0.02$ & $16.0\pm1.8$ & \nodata \\
I3 & 1 & \nodata & $0.12\pm0.02$ & $11.2\pm1.9$ & $1993.3\pm0.6$ \\
   & 2 & $0.03\pm0.01$ & $0.13\pm0.02$ & $11.7\pm1.4$ & \nodata \\
I4 & 1 & \nodata & $0.18\pm0.04$ & $16.2\pm3.9$ & $1998.2\pm0.8$ \\
I5 & 1 & \nodata & $0.09\pm0.04$ & $8.7\pm4.1$ & $1999.0\pm1.2$ \\
\enddata
\tablecomments{See text for the definitions of $\dot{\mu}$, $\langle \mu
\rangle$, $\langle \beta_\mathrm{app} \rangle$, and $T_0$.}
\tablenotetext{a}{Order of the fitted polynomial.}
\end{deluxetable}

Component I4 seems to show motion that indicates acceleration similar
to I3, but since there are only four data points for I4, we have not
fitted it with a second-order polynomial, which would leave only one
degree of freedom for the fit. By looking at Figure~\ref{motion}, one
can see that I4 should have been present in the jet already at the
epoch 1999.55, but it cannot be unambiguously inserted into the source
model; i.e. adding one more component in the model does not improve
the fit, but rather makes the final fit highly dependent on the chosen
initial model parameters. One possible explanation is that the proper
motion for I4 is faster than indicated by the best-fit line in
Figure~\ref{motion}, and it is not actually present in the jet in
1999.55. A second possibility is that I4 is a trailing shock released
in the wake of the primary superluminal component I3 instead of being
ejected from the core \citep{agu01,gom01}. This alternative is not a
very plausible one, since according to \citet{agu01} trailing shocks
should have a significantly slower speed than the main perturbance.  A
third possible explanation for the non-detection of I4 in 1999.55 is
that either I4 and the core or I4 and I3 are so close to one another
in 1999.55 that they do not appear as separate components in our data.

\begin{figure}
\includegraphics[width=\columnwidth]{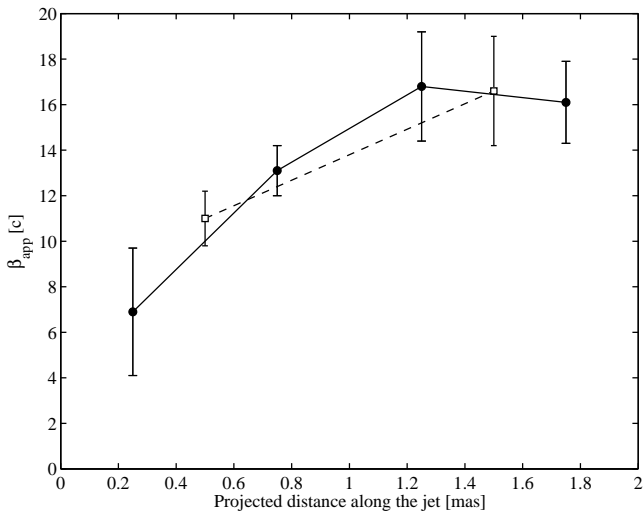}
\caption{Apparent velocity of the components as a function of
projected distance along the jet. The figure shows two different bin
widths: {\it open squares} -- a bin width of 1.0 mas, and {\it filled
circles} -- a bin width of 0.5 mas. \label{beta}}
\end{figure}

As discussed above, we cannot firmly establish that component I3
accelerates along the jet ridge line, although it seems
probable. However, also the other components (apart from I4, which
itself might show acceleration) appear to be systematically faster,
the farther away from the core they lie. We have averaged the
velocities of the individual components over traveled distance bins of
1.0 and 0.5 mas. In averaging, we have taken into account the standard
errors given in Table~\ref{speeds} and the fact that for a given
component in a given bin, the weight has to be decreased, if the first
or the last observation of the component falls into that bin. This is
because naturally we do not have any knowledge about the component
speed before or after our observations. For component I3 we have used
the quadratic fit instead of the linear, since it much better
describes the data. The binned velocities are presented in
Figure~\ref{beta}, which suggests that there is acceleration along the
jet ridge line. When the data is divided in four bins, the first bin
(0--0.5 mas) has an average component velocity of $\approx7\pm3\,c$,
while the rest of the bins (0.5--2.0 mas) have significantly higher
velocities, the maximum being $\approx17\pm3\,c$ for the bin covering
1.0--1.5 mas. For the division in two bins, the change is from
$\approx11\pm1\,c$ in the first bin to $\approx17\pm3\,c$ in the
second. The acceleration takes place in the first strong bend of the
jet, which seems very natural since the apparent velocity depends on
the angle between the local direction of the jet flow and our line of
sight. The acceleration in Figure~\ref{beta} can alternatively be
explained with the component speed changing as a function of ejection
epoch -- a new component (I5) being slower than the old ones
(I1--I2). This, however, does not explain the likely acceleration of
I3. Thus we regard the change in $\beta_{\mathrm{app}}$ along the jet
ridge line as probable although more data is needed for conclusive
evidence.

\citet{kel04} parameterized the $1995-2001$ VLBA 2 cm Survey data of
\objectname{PKS\,2136+141} by fitting Gaussian components to the
brightest features of each epoch in the image plane and derived a
speed of $\beta_{\mathrm{app}}=1.8\pm1.4$, which is much lower than
our estimates. Basically, their approach traces the component I3 over
the time when its proper motion was slow (see Figure~\ref{motion}),
and due to the curved path of I3, the speed measured from the changes
in its radial distance from the core is also lower than its true
angular velocity. A linear least-squares-fit to the radial distances
of I3 in our data over the epochs $1995-2001$ yields
$\beta_{\mathrm{app}}=2.6$, which agrees with the speed given in
\citet{kel04} within the errors.

\subsection{Apparent Half-Opening Angles and Jet Inclination \label{angles}} 
The angle between the local jet direction and our line of sight
affects also the apparent opening angle of the jet. Assuming that the
jet has a conical structure and a filling factor of unity, we have
estimated the local value of the apparent half-opening angle for
component $i$ as $\Psi^{\mathrm{app}}_i = \tan ^{-1} (d_i/l_i)$, where
$l_i$ is the distance of the component $i$ from the core measured
along the jet ridge line, and $d_i=a_i/2+\Delta r_i$, with $a_i$ being
the size of the elliptical component projected on the line
perpendicular to the jet and $\Delta r_i$ being the normal distance of
the component from the jet ridge line.

\begin{figure}
\includegraphics[width=\columnwidth]{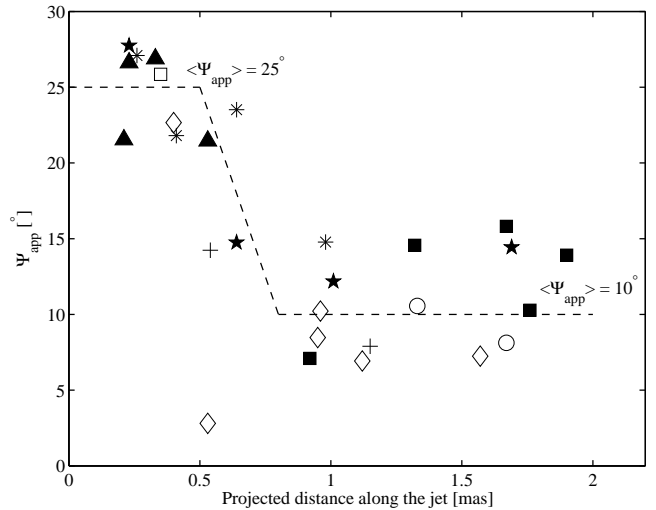}
\caption{Apparent half-opening angle $\Psi_{\mathrm{app}}$ of the
components as a function of projected distance along the jet. The
symbols for components I1-I6 are the same as used in Figs. 6 and
7. {\it Stars} correspond to components of the 22 GHz data and {\it
asterisks} are for the 43 GHz components. The dashed line shows the
average $\Psi_{\mathrm{app}}$ in two regions: 1) the innermost 0.5
mas, and 2) between 0.8--2.0 mas.}
\label{psi}
\end{figure}

Figure~\ref{psi} shows the apparent half-opening angle for components
I1-I6 from 15 GHz data and for unnamed components from 22 and 43 GHz
data as a function of traveled distance along the ridge line. The
components with zero axial ratio have been excluded from the figure.
Naturally, $\Psi_{\mathrm{app}}$ is very uncertain for a single
component, but Figure~\ref{psi} clearly shows that there is a change
in the average $\Psi_{\mathrm{app}}$ between the first $\sim 0.5$ mas
and $0.8-2.0$ mas, and the change seems to correspond to the first
large bend in the jet. Within the first 0.5 mas, the average apparent
half-opening angle is $\sim 25\degr$ and further down the jet it
decreases to $\sim 10\degr$. A natural explanation for this effect is
that the viewing angle $\theta$ increases in the first apparent bend,
since $\Psi_{\mathrm{app}} \approx \Psi_{\mathrm{int}}/\sin \theta$ if
$\theta$ and the intrinsic half-opening angle $\Psi_{\mathrm{int}}$
are both small. There is a caveat, though. Namely, the large value of
$\Psi_{\mathrm{app}}$ near the core could be due to a resolution
effect: if the positional uncertainties for the components near the
core were $\sim 1/5$ of the beam size, $\Delta r$ of these components
could be $\approx 0.1-0.2$ mas at 15 GHz merely due to the scatter,
and this would result in $\langle \Psi_{\mathrm{app}} \rangle
=25\degr$.  However, there are two reasons, why we think the observed
change in $\Psi_{\mathrm{app}}$ is a true effect and due to a changing
$\theta$. First, the model-fit components from 22 and 43 GHz data show
similar values for $\Psi_{\mathrm{app}}$ near the core as the
components from 15 GHz data, although they have a factor of 1.5--2
smaller positional errors. Secondly, the components near the core are
bright, having fluxes comparable to the core, which makes their
positional uncertainties smaller than the adopted $\sim 1/5$ of the
beam size (see also the previous section for a discussion about the
positional uncertainties). According to \citet{jor05} bright (flux of
the knot $>100$ rms noise level) and compact (size $< 0.1$ mas)
features in the jet have positional uncertainties $\sim 0.01$ mas at 7
mm. Thus we regard the observed change in $\Psi_{\mathrm{app}}$ as a
genuine effect.

The $\beta_{\mathrm{app}}$ of the jet seems to increase after $\sim
0.5$ mas from the core (see Figure~\ref{beta}) while the
$\Psi_{\mathrm{app}}$ decreases. This can be understood, if the angle
between local jet direction and our line of sight within the first
$\sim 0.5$ mas is smaller than the angle $\theta_{SL}=\sin ^{-1}
(1/\Gamma)$, which maximizes $\beta_{\mathrm{app}}$. After 0.5 mas
$\theta$ increases towards the maximal superluminal angle increasing
$\beta_{\mathrm{app}}$ and decreasing $\Psi_{\mathrm{app}}$. Assuming
that the largest velocity in Figure~\ref{beta},
$\beta_{\mathrm{app}}\approx17\,c$, is close to the maximum apparent
velocity (we do not consider the velocity of I1, 25.1\,$c$, very reliable
since it is based on only two data points, and hence, we do not use it
in our estimation of the maximum $\beta_{\mathrm{app}}$), we can
estimate that the jet Lorentz factor $\Gamma \sim 20$. Now, if we
assume a constant Lorentz factor $\Gamma = 20$, we can fit for the
values of the viewing angle and the intrinsic jet half-opening angle
giving the observed superluminal speeds and half-opening angles before
and after the bend. For $\langle \beta_{\mathrm{app}} \rangle \approx
7\,c$, $\langle \Psi_{\mathrm{app}} \rangle \approx 25\degr$ within
the first 0.5 mas from the core and $\langle \beta_{\mathrm{app}}
\rangle \approx 17\,c$, $\langle \Psi_{\mathrm{app}} \rangle \approx
10\degr$ after the bend, we obtain the following best-fit values:
$\theta(0-0.5\,\mathrm{mas}) = 0.6\degr$,
$\theta(1.0-2.0\,\mathrm{mas}) = 1.5\degr$, and
$\Psi_{\mathrm{int}}=0.26\degr$. According to this result, the jet
bends away from our line sight by $\approx0.9\degr$ after the first
0.5 mas from the core. 

Naturally, it is possible that the components exhibit also real
acceleration along the jet in addition to the apparent acceleration
due to the projection effect. Unfortunately, the observations do not
allow us to decide between these cases. Applying Occam's razor, we
consider a constant jet speed in the following discussion.

\section{DISCUSSION \label{disc}}

\subsection{Possible Reasons for Bending}
In this section we discuss the mechanisms capable of producing the
curved structure of \objectname{PKS\,2136+141}. In principle, there
are several possible scenarios for the observed bending in
relativistic jets, but we can rule out some of them in this particular
case on the grounds of our analyzed data.

Projection effects play an important role in this case making the
intrinsic bending angle much less than the observed one. If the angle
between the jet and our line of sight is as small as the analysis in
the previous section indicates, it is possible that, for example,
within 0.5-1.0 mas from the core where the apparent bend is $\sim
90\degr$, the intrinsic bending angle is only $\sim 1-2$ degrees. This
implies that formation of internal shocks in the bend, and their
possibly destructive effect to the collimation of the jet, is not of
great concern here. \citet{men02} calculate an upper limit to the
bending angle of a jet in order not to create a shock wave at the end
of the curvature. Their result for relativistic jets is $\sim
50\degr$, which leaves our estimated intrinsic bending angle for
\objectname{PKS\,2136+141} well within the limits. Therefore, we do
not consider internal shocks formed by bending to restrict the
possible explanations for the curvature in this source.

One of the scenarios we can rule out, is a precessing jet where the
components ejected at different times to different directions move
ballistically, and form an apparently curved locus. Such models have
been used to describe oscillating 'nozzles' observed in some BL Lac
sources \citep{sti03,tat04}. In the case of
\objectname{PKS\,2136+141}, we can reject the precessing ballistic jet
hypothesis, since the individual components do not follow ballistic
trajectories, but rather exhibit streaming motion along a curved path
(see \S \ref{sec_traj}). However, precession of the jet inlet may
still be behind the observed structure if it serves as an initial
perturbation driving a helical Kelvin-Helmholtz normal mode (see \S
\ref{model}).

\citet{hom02} explained the large misalignment between the pc and kpc
scale jets of \objectname{PKS\,1510-089} with a scenario where the jet
is bent after it departs the host galaxy, either by a density gradient
in the transition region or by ram pressure due to the winds in the
intracluster medium. In \objectname{PKS\,2136+141} the bending starts
within 0.5 mas from the core, and because $\theta=0.6_{-0.3}^{+0.2}$
degrees for the inner part of the jet (the error range here refers to
an uncertainty introduced by errors in $\langle \beta_{\mathrm{app}}
\rangle$; see \S~\ref{velocities} and \ref{angles}), the corresponding
deprojected distance is smaller than 0.8 kpc. The whole bending
visible in Figure~\ref{mutka} takes place within about 15 mas from the
core. Given the viewing angles estimated in \S~\ref{angles}, it is
fair to say that the deprojected length of the jet is -- probably
significantly -- less than $\sim 12$ kpc, and the whole observed
bending takes place within that distance from the core. There are no
observations of the host galaxy of \objectname{PKS\,2136+141}, but the
deprojected lengths estimated above can be compared with typical
scalelengths of the elliptical hosts of radio-loud quasars (RLQs),
which are reported by several groups. In their near-infrared study,
\citet{tay96} found low-redshift $(z\sim0.2)$ RLQ hosts to have very
large half-light scalelengths $R_{\mathrm{1/2}}$; from 14.0 to 106.8
kpc with an average of $\sim 30$ kpc. Other studies have yielded
smaller values: e.g. \citet{flo04} used {\it Hubble Space Telescope}
WFPC2 data to study hosts of 17 quasars at $z\sim0.4$ finding $\langle
R_{\mathrm{1/2}} \rangle = 10.2\pm1.8$ kpc for RLQs, and \citet{kot98}
reported $\langle R_{\mathrm{1/2}} \rangle = 13\pm7$ kpc for 12 flat
spectrum radio quasars up to $z=1.0$ in their near-infrared study. In
the framework of hierarchical models of galaxy formation, the hosts of
quasars at high redshift are also expected to be more compact than
their low-redshift counterparts. \citet{fal05} have recently managed
to resolve a quasar host at $z=2.555$ and they report an effective
radius of $7.5\pm3$ kpc. As the deprojected distance $\lesssim0.8$ kpc
indicates, the bending of the jet in \objectname{PKS\,2136+141} starts
well within the host galaxy, and the typical scalelengths listed above
suggest that most of the curved jet is located within the host,
although it is possible that part of it lies in the outskirts of the
host galaxy being susceptible to the density gradient in the
transition region. However, the bending clearly starts inside the
galaxy, and thus requires some other cause.

A collision inside the host galaxy, between the jet and a cloud of
interstellar matter, can change the direction of the flow. For example,
\citet{hom03} have found component C4 in \objectname{3C\,279} to
change its trajectory by $26\degr$ in the plane of the sky, and they
suggest this to be due to a collimation event resulting from an
interaction of the component with the boundary between the outflow and
the interstellar medium. However, the mere fact that the observed
$\Delta$P.A. is larger than 180$\degr$ in our case constrains the
collision scenario since it excludes the situations where the jet is
bent by a single deflection, e.g. from a single massive cloud. In
principle, the observed structure may be due to several consecutive
deflections from a number of clouds, but, as already mentioned in \S
\ref{structure}, this would require there to be at least three
successive deflections having the same sense of rotation in the plane
of the sky. This is unlikely, although not impossible. Future
observations aimed at detecting emission from the jet beyond 20 mas
will be very interesting, since they could tell whether the bending
continues in the larger scale or not. If curvature with the same sense
of rotation as within 15 mas is observed also in the larger scale, the
scenario with multiple deflections from ISM clouds seems highly
unlikely.

\subsection{Helical Streaming Model \label{model}}

Relativistic jets are known to be Kelvin-Helmholtz unstable, and they
can naturally develop helical distortions if an initial seed
perturbation is present at the jet origin. The helical K-H fundamental
mode is capable of displacing the entire jet and consequently
producing large scale helical structures where the plasma streams
along a curved path \citep{har87}. The initial perturbations can have
a random spectrum, and be due to for example jet-cloud interaction, or
they can originate from a periodic variation in the flow direction of
the central engine (precession or orbital motion). These perturbations
can trigger pinch, helical or higher order normal modes propagating
down the jet, and the appearance of this structure depends on the
original wave frequency and amplitude, as well as subsequent growth or
damping of the modes. Since the individual components in
\objectname{PKS\,2136+141} do not follow ballistic trajectories, but
rather seem to stream along a helical path, we consider a helical K-H
fundamental mode to be a possible explanation for the curious
appearance of this source. For simplicity, in the following discussion
we limit ourselves to a purely hydrodynamical case and do not consider
the effect of magnetic fields for the growth of K-H modes, nor discuss
the current-driven instabilities, which can produce helical patterns
in Poynting-flux dominated jets \citep{nak04}.

The appearance of the jet in \objectname{PKS\,2136+141} already suggests
some properties of the wave. The fact that components are observed to
follow a nearly stationary helix implies that the wave frequency of
the helical twist, $\omega$, is much below the ``resonant'' (or
maximally unstable) frequency $\omega^*$, which corresponds to the
frequency of the fastest growing helical wave \citep{har87}. Such a
low frequency, long wavelength helical twist suggests that the wave is
driven by a periodic perturbation at the jet base. If the central
source induced white-noise-like perturbations, we would expect to see
structures corresponding to the fastest growing frequency,
i.e. $\omega^*$ \citep{har94}.

\begin{figure*}
\includegraphics[width=\textwidth]{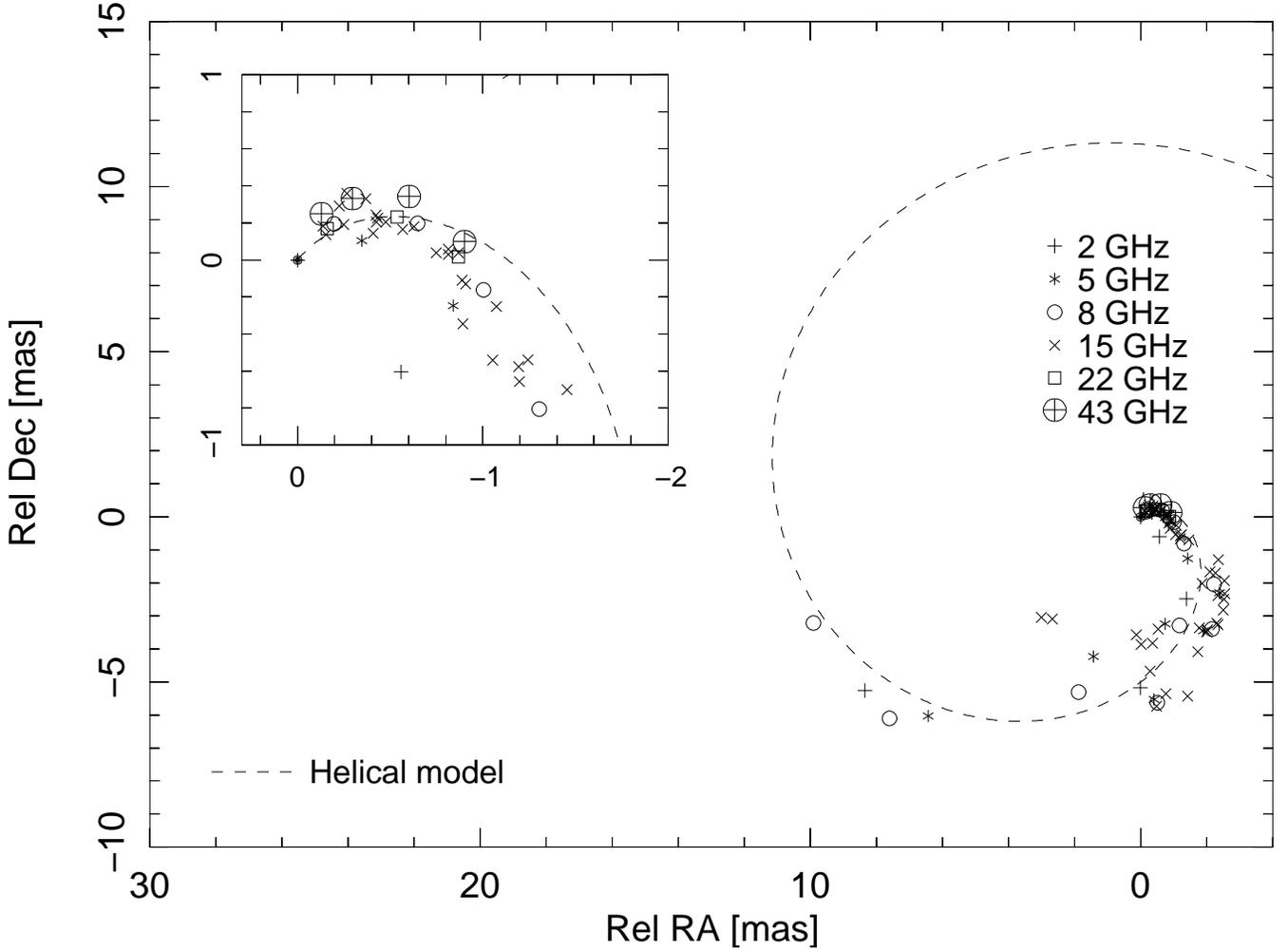}
\caption{Best-fit trajectory and component centroid positions at
different frequencies for the model describing a low-frequency helical
fundamental mode of Kelvin-Helmholtz instability.}
\label{fit}
\end{figure*}

If isothermal jet expansion without gradients in the jet speed or in
the ratio of the jet and external medium sound speeds is assumed, the
wave speed, $\beta_w$ (in units of $c$), in the low frequency limit
remains constant as the jet expands \citep{har03}. Since the intrinsic
wavelength, $\lambda$, for a given $\omega$ varies proportional to the
the wave speed, it can be also assumed to remain constant along the
jet (as long as $\omega \ll \omega^*$). Applying this assumption, we
have fitted a simple helical twist to the observed positions of the
VLBI components. The helical twist is specified in cylindrical
coordinates with $z$ along the axis of helix, by an amplitude $A$ in
the radial direction, and by a phase angle $\phi$ given by
\begin{equation} \label{eq_phi}
\phi = 2 \pi h \int_{z_1}^z \frac{dz}{\lambda(z)} + \phi_1,
\end{equation}
where $h$ is the handedness of the helix ($-1$ for right-handed and
$+1$ for left-handed), and $\phi_1=\phi(z_1)$. With constant
$\lambda$, the integral in equation~(\ref{eq_phi}) becomes trivial and
$\phi \propto z$. The amplitude growth is assumed to be conical: $A =
z \tan \psi_c$, where $\psi_c$ is the opening angle of the helix cone.
To describe the orientation of the helix in the sky we use two angles:
$i$ is the angle between the axis of the helix and our line of sight
and $\chi$ is the position angle of the axis of the helix projected on
the plane of the sky. There are altogether five parameters in the
model since the handedness can be fixed to $h=+1$ by simply looking at
the images.

\begin{figure}
\includegraphics[width=\columnwidth]{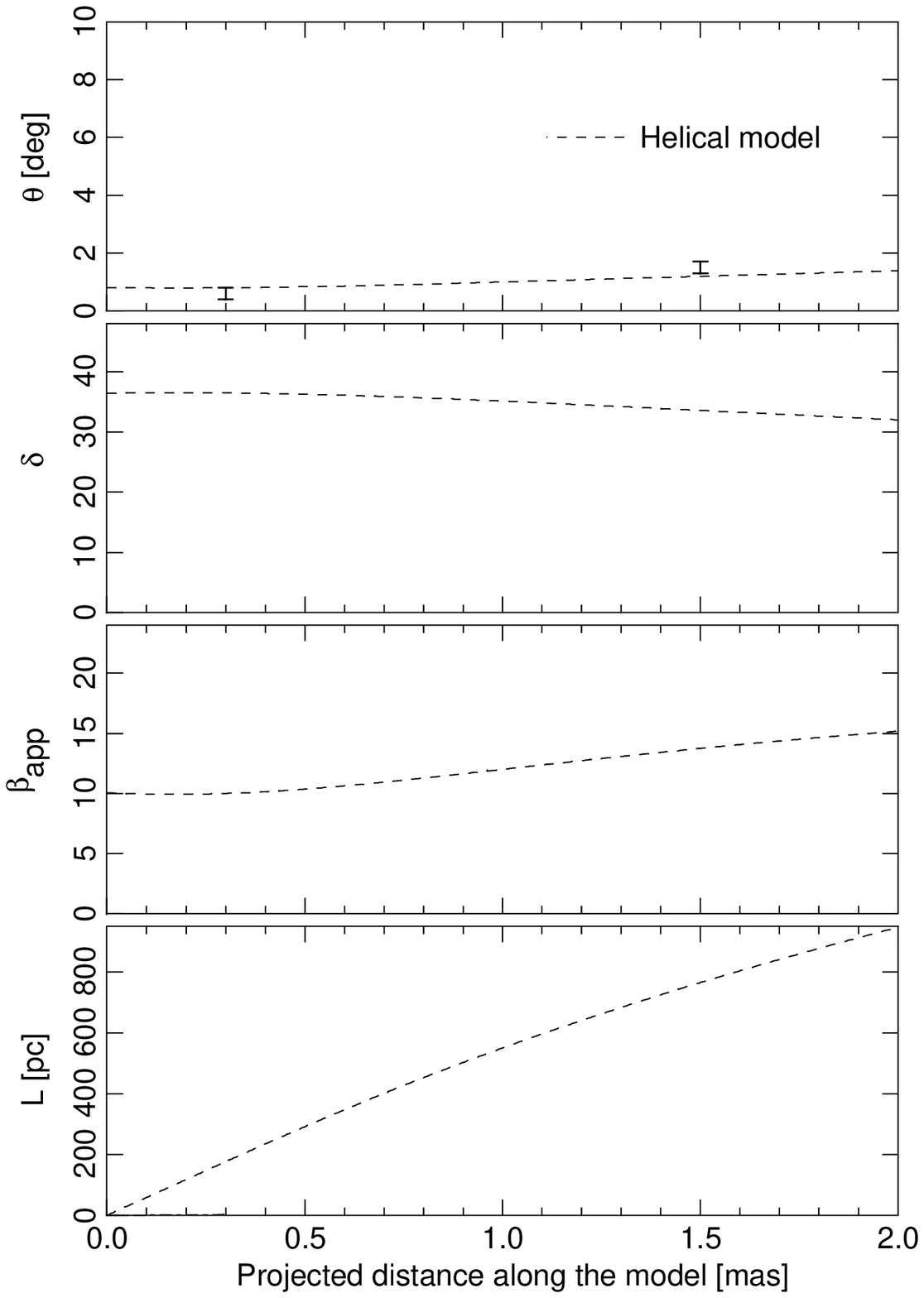}
\caption{Viewing angle {\it (top)}, Doppler factor {\it (second from
the top)}, apparent component speed {\it (third from the top)} and
deprojected distance {\it (bottom)} for a jet with $\Gamma$ = 20 as a
function of projected distance from the core along a helical model
(see Figure~\ref{fit}). The two small vertical bars in the top panel
show the viewing angle constraints obtained from the source kinematics
and used in the modeling.}
\label{model_angles}
\end{figure}

A computer program was written to fit the positions of the VLBI
components with different models expected to explain the shape of the
jet and the kinematics of individual components. In the program, the
cumulative sum of perpendicular projected image-plane distances
between a three-dimensional model and the component centroids from the
VLBI observations was minimized using a differential evolution
\citep{sto97} algorithm. This evolutionary algorithm was chosen
because of its good performance with non-linear and multimodal
problems. All the observed components (and all frequencies) were used
in a single model-fit, because the components seem to follow a common
path, i.e. the helical structure appears to persist for longer than
the component propagation times. In addition to component positions on
the plane of the sky, the model was constrained by the viewing angles
determined in \S \ref{angles}. The angle $\theta$ between the normal
to the jet's cross section and our line-of-sight was evaluated along
the model, and it was required to be compatible with the viewing
angles determined in \S \ref{angles}. The following limits for
$\theta$ were used: at 0.3 mas from the core $\theta$ is between
$0.4-0.8 \degr$, and at 1.5 mas from the core $\theta$ is between
$1.3-1.7\degr$. If the model exceeded these limits, the cost function
in the fitting algorithm was multiplied with a second-order function
normalized to the width of the allowed range.

\begin{deluxetable}{lcc}
\tablewidth{0pt}
\tablecaption{Best-fit Parameters of the Helical
Streaming Model \label{dri_model}}
\tablecolumns{3}
\tablehead{\colhead{Parameter} & \colhead{Symbol} & \colhead{Value}}
\startdata
Initial helical wavelength [mas] & $\lambda_1$ & 776 \\
Half-opening angle of the helix cone [deg] & $\psi_c$ & 1.0 \\
Initial phase angle [deg] & $\phi_1$ & -69 \\
Viewing angle to cone axis [deg] & $i$ & 0.3 \\
Sky position angle of cone axis [deg] & $\chi$ & -158 \\
Helix handedness & $h$ & +1 \\
\enddata
\end{deluxetable}

The best-fit trajectory of the model is presented in Figure~\ref{fit},
and Figure~\ref{model_angles} shows the angle between the local jet
direction and our line of sight, the Doppler factor, and the apparent
velocity as a function of distance along the jet, which have been
calculated by assuming $\Gamma=20$. As is clear from these figures,
the simple helical twist with constant wavelength along the jet gives
a good fit to the data. The best-fit parameters of the model are
listed in Table~\ref{dri_model}. In the best-fit model, we are looking
straight into the cone of the helix ($i=0.3\degr$), which has a small
half-opening angle of $1.0\degr$, meaning that the orientation of the
helix is a very lucky coincidence. The fitted helical wavelength of
the perturbation is 776 mas, corresponding to 6.4 kpc. This still
needs to be corrected for a combined effect of the geometry and the
possibly relativistic wave speed. The true intrinsic helical
wavelength is given by $\lambda_\mathrm{int}=\lambda_\mathrm{model}
(1-\beta_w \cos i)$. Some constraint for $\beta_w$ can be derived from
the fact that we do not see any systematic change in the helical
trajectory during 8.7 years, i.e. the change of the trajectory is less
than the components' positional uncertainty in our observations. This
gives an upper limit on the apparent wave speed, $\beta_w^\mathrm{app}
< 1.1\,c$. As we look into the helix cone, the appropriate viewing
angle for the wave motion is between $\psi_c - i$ and $\psi_c + i$,
which gives $\beta_w < 0.989$ and $\lambda_\mathrm{int} >
0.01\lambda_\mathrm{model}$. These limits are purely due to the
uncertainty in the positions of the VLBI components ($\sim 1/5$ of the
beam size) and it is likely that the true wave speed is much slower
and the intrinsic wavelength is closer to
$\lambda_\mathrm{model}$. Further observations, e.g. within MOJAVE
program, should give a tighter constraint for $\beta_w$.  In the low
frequency limit, the wave speed is
\begin{equation} \label{beta_w}
\beta_w = \frac{\Gamma^2\eta}{1+\Gamma^2\eta}\beta_j,
\end{equation}
where $\beta_j=\sqrt{1-\Gamma^{-2}}$ is the flow velocity, and
$\eta=(a_x/a_j)^2$, with $a_j$ and $a_x$ being the sound speeds in the
jet and in the external medium, respectively \citep{har03}. Assuming
$\Gamma=20$ and applying the upper limit of $\beta_w$, we get a limit
on the ratio of sound speeds: $a_x/a_j < 0.5$.

We would like to have an estimate of the wave frequency of the
observed helical twist since it could possibly tell us about the
origin of the periodic perturbation. Unfortunately, we do not know the
wave speed, and hence cannot calculate the frequency. The limiting
case with $\beta_w=0.989$ considered above yields $\omega \sim 1.4
\times 10^{-10}$\,Hz corresponding to a period of $\sim 200$\,yr,
which is likely much below the actual period. However, some example
values can be calculated for different combinations of $a_x$ and
$a_j$. Let us first assume an ultra-relativistic jet with $a_j =
c/\sqrt{3}$ and $\Gamma = 20$. Now, using equation~(\ref{beta_w}) we
can calculate $\beta_w$ and consequently $\omega$ for different values
of $a_x$. For example, \citet{con95} use $a_x \approx 400$ km/s in
their study of \objectname{Mrk\,501} on basis of X-ray observations
and theoretical modeling of giant elliptical galaxies, but this value
refers to an average sound speed in the central regions of an
elliptical galaxy. Around the relativistic jet, there may be a hot
wind or cocoon, where the sound speed is much higher, being a
significant fraction of the light speed. For instance, \citet{har05}
estimate that the sound speed immediately outside the jet in the radio
galaxy \objectname{3C\,120} is $a_x \gtrsim 0.1\,c$. For these two
values, $a_x = 0.001c$ and $a_x = 0.1c$, we get $\omega \sim
2\times10^{-15}$\,Hz and $\omega \sim 2\times10^{-13}$\,Hz,
respectively. The corresponding driving periods are $P_d \approx
10^7$\,yr and $P_d \approx 10^5$\,yr. If $a_j$ is less than
$c/\sqrt{3}$ of the ultra-relativistic case, the frequencies will be
higher and corresponding periods shorter.

The helical streaming model presented above describes only one normal
mode and it does not explain why the components in the 15 GHz
monitoring data cluster in groups, i.e. why there seem to be certain
parts in the jet where the flow becomes visible (see \S~3.1). However,
this might be explained if there are other, short wavelength,
instability modes present in the jet in addition to the externally
driven mode. Our current data set does not allow to test this
hypothesis, but some hints of another instability mode may be present
in Figure~\ref{fit} where the components look like they are
``oscillating'' about the best-fit trajectory.

\section{CONCLUSIONS \label{conc}}

We have presented multi-frequency VLBI data revealing a strongly
curved jet in the gigahertz-peaked spectrum quasar
\objectname{PKS\,2136+141}. The observations show a $210\degr$ change
of the jet position angle, which is, to our knowledge, the largest
ever observed $\Delta P.A.$ in an astrophysical jet. The jet
appearance is highly reminiscent of a helix with the axis of the helix
cone oriented towards our line of sight.

Eight years of monitoring with the VLBA at 15 GHz show several
components moving down the jet with clearly non-ballistic
trajectories, which excludes the precessing ballistic jet model from
the list of possible explanations for the helical structure in
\objectname{PKS\,2136+141}. Instead, the individual components are
streaming along a curved trajectory. The estimated ejection epochs of
the components are coincident with two major total flux density
outbursts in 1990's, with three components being associated with both
outbursts. This may suggest that a single flare-event is associated
with complicated structural changes in the jet, possibly involving
multiple shocks.

Most of the components have apparent velocities in the range of
$8-17\,c$. One component even shows $25\,c$, but this high speed is
based on only two data points, and therefore it is unreliable. We find
evidence suggesting that $\beta_\mathrm{app}$ increases after the
first 0.5 mas from the core, a distance corresponding to a strong
bend. Since the apparent jet opening angle also changes at this point,
we suggest that the angle between the local jet direction and our line
of sight increases from $\approx 0.6\degr$ mas within first 0.5 mas to
$\approx 1.5\degr$ at distances between 1 and 2 mas.

We fit the observed jet trajectory with a model describing a jet
displaced by a helical fundamental mode of Kelvin-Helmholtz
instability. Our observations suggest that the wave is at the low
frequency limit (relative to the ``resonant'' frequency of the jet)
and has a nearly constant wave speed and wavelength along the
jet. This favors a periodic perturbation driven into the jet. The
source of the perturbation could be e.g. jet precession or orbital
motion of a supermassive binary black hole. Our present data does not
allow us to reliably calculate the driving period of the K-H wave, and
hence we cannot further discuss the perturbation's origin. Follow-up
observations of the jet at angular scales larger than 15 mas, as well
as further monitoring of the jet kinematics, will probably shed light
on this question. In the best-fit model, the helix lies on the surface
of the cone with a half-opening angle of $1.0\degr$, and the angle
between our line of sight and the axis of the cone is only $0.3\degr$,
i.e. we are looking right into the helix cone.

\acknowledgments The authors thank the referee, Philip Hardee, for his
suggestions, which significantly improved the helical streaming model
section. This work was partly supported by the Finnish Cultural
Foundation (TS), by the Japan Society for the Promotion of Science
(KW), and by the Academy of Finland grants 74886 and 210338.  MK was
supported through a stipend from the International Max Planck Research
School for Radio and Infrared Astronomy at the University of Bonn. We
gratefully acknowledge the VSOP Project, which is led by the Institute
of Space and Astronautical Science of the Japan Aerospace Exploration
Agency, in cooperation with many organizations and radio telescopes
around the world and H.~Hirabayashi in particular for allowing us to
use unpublished data. We also thank P.~Edwards for help in amplitude
calibration of the HALCA dataset. We made use of data obtained as part
of the VLBA 2\,cm Survey. The VLBA is a facility of the National Radio
Astronomy Observatory, operated by Associated Universities Inc., under
cooperative agreement with the U.S. National Science Foundation. UMRAO
is supported in part by funds from the NSF and from the University of
Michigan Department of Astronomy.

\end{document}